\documentclass[aps,prc,groupedaddress,onecolumn,floatfix]{revtex4}
\usepackage{amsmath}
\usepackage{amssymb}
\usepackage{amsfonts}
\usepackage{amscd}
\usepackage{bm}
\usepackage{graphicx}
\usepackage{color}
\def\beq{\begin{equation}}
\def\eeq{\end{equation}}

\begin{document}

\title{Euclidean
formulation of relativistic quantum mechanics of N particles.} 

\author{Gohin Shaikh Samad}
\affiliation{Department of Mathematics\\
The University of
Iowa\\ Iowa City, IA 52242, USA}

\author{W.~N.~Polyzou}
\email{polyzou@uiowa.edu}
\thanks{This work supported by the U.S. Department of Energy,
  Office of Science, Grant \#DE-SC16457}
\affiliation{Department of Physics and Astronomy\\ The University of
Iowa\\ Iowa City, IA 52242, USA}

\date{\today}

\begin{abstract}

  A Euclidean formulation of relativistic quantum mechanics for
  systems of a finite number of degrees of freedom is
  discussed. Relativistic treatments of quantum theory are needed to
  study hadronic systems at sub-hadronic distance scales.  While
  direct interaction approaches to relativistic quantum mechanics have
  proved to be useful, they have two disadvantages. One is that
  cluster properties are difficult to realize for systems of more than
  two particles. The second is that the relation to quantum field
  theories is indirect.  Euclidean formulations of relativistic
  quantum mechanics provide an alternative representation that does
  not have these difficulties.  More surprising, the theory can be
  formulated entirely in the Euclidean representation without the need
  for analytic continuation.  In this work a Euclidean representation
  of a relativistic $N$-particle system is discussed.  Kernels for
  systems of N free particles of any spin are given and shown to be
  reflection positive.  Explicit formulas for generators of the Poincar\'e
  group for any
  spin are constructed and shown to be self-adjoint on the Euclidean
  representation of the Hilbert space.  The structure of correlations
  that preserve both the Euclidean covariance and reflection
  positivity is discussed.

\end{abstract}

\maketitle 

\section{Introduction}

Relativistic quantum mechanical models of systems with a finite number
of degrees of freedom are useful for modeling strongly interacting
systems because they can be solved numerically with controlled errors
and can be applied consistently in both the lab frame and center of
momentum frame.  This paper discusses a Euclidean covariant
representation of relativistic quantum mechanics for systems with a
finite number of degrees of freedom.  The Euclidean representation
overcomes some of the difficulties with the direct construction of
Poincar\'e generators on a multi-particle Hilbert space.  The two key
challenges of the direct construction are constructing generators
satisfying cluster properties in all inertial coordinate systems for
systems of more than three particles and the absence of a direct
relation between the model interactions and an underlying quantum
field theory.

The motivation for exploring the Euclidean formulation is that
it provides a representation of a relativistic quantum theory that has
a direct connection to quantum field theory and easily satisfies
cluster properties.  While the same is formally true of Minkowski
representations of quantum field theory, most non-perturbative
computations are based on relations among time-ordered vacuum
expectation values of fields, while the Hilbert space structure, which
is associated with the Wightman functions, is simply assumed.  When
truncations are involved it is not automatic that solutions of
Schwinger-Dyson equations with phenomenological input are consistent
with the probabilistic interpretation of quantum theory.  Both
Wightman functions and time-ordered Green functions are related to
Euclidean Green functions by different analytic continuations.  The
Euclidean Green's functions satisfy Euclidean versions of the
Schwinger-Dyson equations and at the same time are directly related to
the Hilbert space structure of the field theory.  An appealing feature
of the Euclidean axioms is that the locality axiom is logically
independent of the other axioms, so it can be relaxed (which is
necessary for models of a finite number of degrees of freedom) without
violating relativistic invariance, the spectral condition, cluster
properties, and the Hilbert space representation of the theory.
A second appealing feature is that an analytic continuation is not
necessary to compute the Hilbert space inner product of the physical
quantum theory.  While the formulation of the dynamics discussed in
this work is still phenomenological, the phenomenological kernels are
in principle models of the exact Euclidean Green functions of the
field theory, so they can be constrained by field theory based
phenomenology.

In a quantum theory relativistic invariance means that quantum
observables, which are probabilities, expectation values and ensemble
averages, have the same value for equivalent experiments that are
performed in different inertial coordinate systems.  This means that
experiments performed in an isolated system cannot be used to
distinguish inertial coordinate systems.  In special relativity
different inertial coordinate systems are related by the subgroup of
Poincar\'e group connected to the identity.  In 1939 E. P. Wigner
\cite{Wigner:1939cj} showed that a necessary and sufficient condition
for a quantum system to be relativistically invariant is that vectors
representing equivalent quantum states in different inertial
coordinate systems are related by a unitary ray representation of this
subgroup on the Hilbert space of the quantum theory.  V. Bargmann \cite{Bargmann:1954gh}
showed that this can be replaced by a single-valued representation of
$SL(2,\mathbf{C})$, which is the covering group of the Lorentz group.

Relativistically invariant quantum theories are needed to study
physics on distance scales that are small enough to be sensitive to
the internal structure of a nucleon.  This is because in order to get
wavelengths short enough to resolve the internal structure of a
nucleon it is necessary to transfer momentum to the nucleon that is
comparable to or larger than its mass scale.

The direct approach for modeling relativistic systems is to construct
explicit expressions for the Poincar\'e generators of the interacting
system on a many-particle Hilbert space.  Formally the
Hilbert space is a direct sum of tensor products of irreducible
representation spaces of the Poincar\'e group, representing the
particle content of the system.  Phenomenological interactions are
added to the non-interacting Poincar\'e generators in a manner that
preserves the commutation relations, cluster properties and the
spectral condition.  This is the referred to as the direct interaction
representation.  This representation shares many of the computational
advantages of non-relativistic quantum mechanics.  One problem is that
the interactions are generally phenomenological and representation
dependent, which makes them difficult to constrain by a more
fundamental theory.  In addition, satisfying cluster properties in all
inertial coordinate systems puts strong constraints on the structure
of the interactions.  Satisfying these constraints presents
computational challenges that have not been realized in applications
\cite{Sokolov:1977}\cite{Coester:1982vt}\cite{Keister:1991sb}.

Strong interactions are studied using lattice methods, which break
relativistic invariance, Schwinger-Dyson equations, which are infinite
systems of non-linear equations Euclidean Green functions, and
relativistic quantum mechanical models, which are more
phenomenological and not directly related to an underlying quantum
field theory.  Both lattice calculations and Schwinger-Dyson
calculation are normally formulated in a Euclidean representation.
The purpose of this work is to formulate a class of relativistic
quantum mechanical model that have many of the properties of direct
interaction relativistic quantum models but have a more direct
connection to lattice and Schwinger-Dyson methods.  The Euclidean
formulation facilitates the relation to these other methods.  The
formulation of relativistic quantum mechanics that will be discussed
in this paper is motivated by the Euclidean reconstruction theorem of
axiomatic quantum field theory.

Euclidean formulations of quantum field theory were first advocated by
J. Schwinger \cite{Schwinger:pna}\cite{Schwinger:1959zz} who used the
spectral condition in time-ordered Green's functions to establish the
existence of an analytic continuation to imaginary times.
Independently, axiomatic treatments of quantum field theory
\cite{Wightman:1980}\cite{jost} led to an understanding of the
analytic properties of vacuum expectation values of products of
fields, based on the spectral condition, Lorentz covariance and locality..  The Euclidean approach
to quantum field theory was advocated by K. Symanzik
\cite{Symanzik:1966}\cite{Symanzik:1968zz}, and developed by E. Nelson
\cite{Nelson:1973}.  K. Osterwalder and R. Schrader
\cite{Osterwalder:1973dx}\cite{Osterwalder:1974tc} identified
properties of Euclidean covariant distributions that are sufficient to
reconstruct a relativistic quantum field theory.  Two observations
that are implicit in the work of Osterwalder and Schrader are (1)
that an explicit analytic continuation is not necessary to construct a
relativistic quantum theory and (2) the reconstruction of a
relativistic quantum theory is not limited to local field theories.
The Euclidean formulation of relativistic quantum mechanics
presented in this work is motivated by these two observations.

An attractive feature of the Euclidean approach is that both the
time-ordered Green functions and Wightman functions can be extracted
from the Euclidean Green functions using different analytic
continuations. This means the Euclidean Green functions satisfy
Euclidean Schwinger-Dyson equations while at the same time they can be
used to construct the Hilbert space inner product of the underlying
quantum theory.  This provides a means to constrain the Hilbert space
formulations of the theory from a Lagrangian based dynamics.  While
this formulation is intended to be phenomenological, it is designed so there
is a formal relation to an underlying quantum field theory.

Aspects of this program have been discussed elsewhere \cite{Kopp:2011vv}\cite{Polyzou:2013nga}
\cite{Aiello:2015jgc}\cite{polyzou:2019}.  The
purpose of this paper is derive explicit expressions for the
Poincar\'e generators with spin and discuss the structure of Euclidean
covariant reflection positive distributions with spin.

This paper is organized as follows.  Notation is introduced in section
2. The relation between the complex Lorentz group and the complex four
dimensional orthogonal group is discussed.  This is central to the
relation between the Euclidean and Lorentz covariant representations
of the theory.  Section 3 discusses positive mass irreducible
representations of the Poincar\'e group.  These are used to construct
equivalent Lorentz and Euclidean covariant representations for massive
particles in section 4.  The relation of Euclidean covariance to
Lorentz covariance is discussed in section 5.  Section 6 contains the
explicit formulas for the Poincar\'e generators with spin.  They
are shown to satisfy the Poincar\'e commutation
relations and be Hermitian on the Euclidean representation of the Hilbert space.
The Generalization to systems of free particles is discussed in section 7.
The inclusion of dynamics and the structure of
dynamical reflection positive Euclidean covariant kernels is
examined in section 8.  Section 9 shows the self-adjointness of the Hamiltonian
and boost generators in the Euclidean representation. Section 10
contains as summary and concluding remarks.  The appendix gives 
space-time representations of reflection positive
Euclidean covariant kernels with different spins.  

\section{Background}

The Poincar\'e group is the group of space-time transformations that
relate different inertial reference frames in the theory of special
relativity.  It is the symmetry group that preserves the proper time
$\tau_{ab}$, or proper distance, $d_{ab}$, between any two events with
space-time coordinates $x_a^{\mu}$ and $x_b^{\mu}$:
\beq
-\tau_{ab}^2=d_{ab}^2=\eta_{\mu\nu}(x_a-x_b)^{\mu}(x_a-x_b)^{\nu}, 
\label{b.1}
\eeq
where $\eta_{11}=\eta_{22}=\eta_{33}=-\eta_{00}=1$, $\eta_{\mu\nu}=0$ for
$\mu \not=\nu$ is the Minkowski metric tensor.  Repeated indices
are assumed to be summed.
The most general point transformation,
$x'^\mu=f^\mu(x)$ satisfying (\ref{b.1}) has the form
\beq
x^{\mu}\to x'^{\mu}=\Lambda^{\mu}{}_{\nu} x^\nu+a^\mu 
\label{b.2}
\eeq
where $\Lambda^{\mu}{}_{\nu}$ is a Lorentz transformation satisfying 
\[
\eta_{\mu\nu}=\Lambda^{\alpha}{}_\mu\eta_{\alpha\beta} \Lambda^{\beta}{}_\nu
\]
or in matrix form 
\beq
\eta=\Lambda^t\eta\Lambda .
\label{b.3}
\eeq

The full Poincar\'e group contains discrete transformations that are
not associated with special relativity.  
Equation (\ref{b.3}) implies that
\beq
\mbox{det}(\Lambda)^2=1 \qquad \mbox{and} \qquad (\Lambda_0^0)^2=
1+\sum_{i}(\Lambda_i^0)^2 .
\label{b.4}
\eeq
This means that the Lorentz group can be decomposed
into four topologically disconnected components
\begin{itemize}
\item 
$\mbox{det}(\Lambda)=1, \qquad(\Lambda_0^0) \geq 1$; includes identity
\item
$\mbox{det}(\Lambda)=-1, \qquad(\Lambda_0^0) \geq 1$; includes space reflection
\item
$\mbox{det}(\Lambda)=-1, \qquad (\Lambda_0^0) \leq-1 $; includes time reversal
\item
$\mbox{det}(\Lambda)=1, \qquad(\Lambda_0^0) \leq -1$; includes space-time reversal .
\end{itemize}
Since the discrete symmetries of space reflection and time reversal
are not symmetries of the weak interaction, the symmetry group
associated with special relativity is normally considered to be the
subgroup of the Poincar\'e group that is continuously connected to the
identity.

The relation between the Lorentz group and the
four-dimensional orthogonal group is central to
the development of the Euclidean formulation.  
The relation is illustrated by
representing Minkowski, $x^{\mu}$, and Euclidean, $x_e^{\mu}$, four
vectors as $2\times 2$ matrices:
\beq
X_m=x^\mu\sigma_\mu=
\begin{pmatrix}
x^0+x^3 &x^1-ix^2  \\
x^1+ix^2 &x^0-x^3 
\end{pmatrix}\; \qquad  x^\mu=\frac{1}{2}\mbox{Tr}(X\sigma_\mu)
\label{b.5}
\eeq
\beq
X_e=x_e^\mu\sigma_{e\mu}=
\begin{pmatrix}
ix_e^0+x^3 &x^1-ix^2  \\
x^1+ix^2 &ix_e^0-x^3 
\end{pmatrix}\; \qquad  x_e^\mu=\frac{1}{2}\mbox{Tr}(X_e\sigma^{\dagger}_{e\mu}).
\label{b.6}
\eeq
where $\sigma_i=\sigma_{ei}$ are the Pauli matrices, $\sigma_0$ is the identity and
$\sigma_{e0}= i \sigma_0$.  The determinants of these matrices are
related to the Minkowski and Euclidean line elements respectively:
\beq
\mbox{det}(X_m) = (x^0)^2 - \mathbf{x}\cdot \mathbf{x}
\qquad 
\mbox{det}(X_e) = - \left ( (x_e^0)^2 + \mathbf{x}\cdot \mathbf{x}
\right ).
\label{b.7}
\eeq
$X_m$ is Hermitian for real four vectors.
The linear transformations that preserve the determinant and hermiticity of
$X_m$ have the form
\beq
X_m \to X_m' = \pm A X_m A^{\dagger}
\qquad \mbox{det} (A)=1.
\label{b.8}
\eeq
The (-) sign represents a space-time reflection, which is not
considered part of the symmetry group of special relativity.  The
group of complex $2 \times 2$ matrices with $\mbox{det}(A)=1$ is
$SL(2,\mathbb{C})$.  Similarly linear transformations corresponding to
real four-dimensional orthogonal transformations in the $2 \times 2$
matrix representation have the general form
\beq
X_e \to X_e' =  \pm A X_e C^t \qquad A,C \in SU(2).
\label{b.9}
\eeq
Transformations of the form
\beq
X_e \to X_e' =  A X_e C^t \qquad 
X_m \to X_m' =  A X_m C^t
\label{b.10}
\eeq
with both $A$ and $C$ in $SL(2,\mathbb{C})$ preserve the Minkowski and
Euclidean line elements respectively,  however they do not preserve
the reality of the four vectors,
\beq
x^{\prime \mu}= {1 \over 2}\mbox{Tr}(X'\sigma_\mu)
\qquad
x_e^{\prime \mu} = {1 \over 2} 
\mbox{Tr}(X'_e\sigma^{\dagger}_{e\mu}).
\eeq
They represent complex Lorentz or orthogonal transformations.  The
corresponding complex Lorentz and orthogonal transformations are
\beq
\Lambda (A,C)^{\mu}{}_{\nu} =
{1 \over 2} \mbox{Tr} (\sigma_{\mu} A \sigma_{\nu} C^{t})
\qquad
O (A,C)^{\mu}{}_{\nu} =
{1 \over 2} \mbox{Tr} (\sigma_{e\mu}^{\dagger}  A \sigma_{e\nu}  C^{t}).
\label{b.12}
\eeq
This shows that the covering group of both the complex Lorentz and
complex orthogonal group is $SL(2,\mathbb{C}) \times
SL(2,\mathbb{C})$.  These are double covers because $A,C \to -A,-C$
result in the same transformation.  For $C=A^*$ eq. (\ref{b.12})
relates the real Lorentz group to a subgroup of the complex orthogonal
group; similarly for $A$ and $C$ unitary eq. (\ref{b.12}) relates the
real orthogonal group to a subgroup of the complex Poincar\'e group.
The relation that will be exploited in this work is that Euclidean
rotations that involve a space and the Euclidean time coordinate can
be identified with Lorentz boosts with complex rapidity.

For the full Poincar\'e group it is necessary to include translations.
Euclidean time translations by $\tau$ are identified with Minkowski
time translations with $t=-i\tau$. 

\section{Unitary representations of the Poincar\'e group}

Any unitary representation of the Poincar\'e group satisfying the
spectral condition ($m>0$) can be decomposed into a direct integral of
positive mass irreducible representations.  The starting point of this
work is to construct positive-mass irreducible unitary representations
of the Poincar\'e group and use them to construct the corresponding
Euclidean representations.  Reflection positive kernels for each
irreducible representation result from this construction.  Since
many-particle Hilbert spaces are tensor products of single-particle
spaces and dynamical unitary representations of the Poincar\'e group
can be decomposed into direct integrals of irreducible representation
spaces, this construction provides a framework for constructing
reflection positive kernels for different physical systems.

$2 \times 2$ matrix representation of four vectors are used in this section.
Poincar\'e group elements are replaced by $(A,Y)$
where A is a $SL(2,\mathbb{C})$ matrix and $Y$ is a
$2 \times 2$ Hermitian matrix representing a translation.  In the
$2 \times 2$ representation Poincar\'e transformations continuously
connected to the identity have the form
\beq
X' = A X
A^{\dagger} + Y
\label{c.1}
\eeq
where the group multiplication law is
\beq
(A_2 , Y_2) (A_1 , Y_1)= (A_2 A_1 , A_2
Y_1 A_2^{\dagger} + Y_2).  
\label{c.2}\eeq
Four vector representations of these equations are
\beq
x^{\mu \prime} = \Lambda^{\mu}{}_{\nu} x^{\nu} + y^{\mu}
\label{c.3}
\eeq
\beq
(\Lambda_{12}^{\mu}{}_{\nu},y_{12}^{\mu} ) =
(\Lambda_2^{\mu}{}_{\alpha} \Lambda_1^{\alpha}{}_{\nu},
\Lambda_2^{\mu}{}_{\alpha}y_1^{\alpha} + y_2^{\mu}),
\label{c.4}
\eeq
where the
four vector representations are related to 
the $2\times 2$ representations by 
\beq
y^{\mu}:= {1 \over 2} \mbox{Tr}(\sigma_{\mu}Y) 
\qquad 
\Lambda^{\mu}{}_{\nu} := {1 \over 2} \mbox{Tr}(\sigma_{\mu}A\sigma_{\nu} A^{\dagger}).
\label{c.5}
\eeq
$SL(2,\mathbb{C})$ is a six parameter group.  It has six independent
one-parameter subgroups
\beq
A_{r \hat{\pmb{\theta}}}({\theta})= e^{{i\theta  \over 2 }\bm{\sigma}\cdot \bm{\hat{\theta}}} = A_r (\pmb{\theta})
\qquad
A_{b \hat{\pmb{\rho}}}(\rho) = e^{{\rho \over 2 }\bm{\sigma}\cdot \bm{\hat{\rho}}} = A_b (\pmb{\rho})
\label{c.6}
\eeq
corresponding to rotations about three different axes and rotationless
Lorentz boosts in three different directions.  In these expressions
$\hat{\bm{\theta}}$ represents the axis and $\theta$
represents angle of a rotation while
$\hat{\bm{\rho}}$ represents the direction of a
rotationless boost and $\rho$ represents
the rapidity of a rotationless boost.
The polar decomposition theorem expresses a general $SL(2,\mathbb{C})$ matrix $A$
as a product of a (generalized Melosh) rotation ($R_m=(R_m^{\dagger})^{-1}$ unitary) followed by
rotationless (canonical) boost ($B_c=B_c^{\dagger}$ positive Hermitian):
\beq
A = B_{c}R_{m} 
\label{c.7}
\eeq
where
\beq
B_{c} :=
(A A^{\dagger})^{1/2} = B_c (\bm{\rho})
\qquad 
R_{m} :=
(A A^{\dagger})^{-1/2}A =R_m (\bm{\theta}).
\label{c.8}
\eeq
A unitary representation of inhomogeneous
$SL(2,\mathbb{C})$ is a set of unitary operators
$U(A ,Y)$, labeled by elements of $SL(2,\mathbb{C}$)
satisfying
\beq
U(A_2,Y_2) U(A_1,Y_1)=
U(A_2 A_1,A_2 Y_1 A_2^{\dagger} +Y_2)
\label{c.9}
\eeq
\beq
U(I,0) =I
\label{c.10}
\eeq
\beq
U^{\dagger}(A ,Y) = U^{-1}(A ,Y)=
U (A^{-1} ,-A^{-1} Y (A^{\dagger})^{-1} ) .
\label{c.11}
\eeq
The Poincar\'e group is a 10 parameter group.  Infinitesimal generators
of $U(A,Y)$ are the 10 self-adjoint operators defined by
\beq
H= i {d \over dy^0} U(I,y^0\sigma_0)_{\vert_{y_0=0}}
\label{c.12}
\eeq
\beq
P^j= - i {d \over dy^j} U(I,y^j\sigma_j)_{\vert_{y_j=0}}
\label{c.13}
\eeq
\beq
J^j= - i {d \over d\theta} U(e^{i{\theta\over 2} \sigma_j},0)_{\vert_{\theta=0}}
\label{c.14}
\eeq
\beq
K^j= - i {d \over d\rho} U(e^{{\rho\over 2} \sigma_j},0)_{\vert_{\rho=0}}
\label{c.15}
\eeq
where there is no sum in (\ref{c.13}) over the repeated $j$, and
$j\in \{1,2,3\}$ in (\ref{c.13}-\ref{c.15}).
The group representation property (\ref{c.9}) implies that these generators
satisfy the Poincar\'e commutation relations
\beq
[J^i ,J^j] = i \epsilon_{ijk} J^k
\qquad
[J^i ,P^j] = i \epsilon_{ijk} P^k
\qquad
[J^i ,K^j] = i \epsilon_{ijk} K^k
\label{c.16}
\eeq
\beq
[K^i ,K^j] = -i \epsilon_{ijk} J^k
\qquad
[J^i ,H] = 0
\qquad
[P^i,H] = 0
\label{c.17}
\eeq
\beq
[K^j,H] =i P^j
\qquad
[K^i ,P^j] = i \delta_{ij}H .
\label{c.18}
\eeq

These operators are components of a four vector, $P^{\mu}$, and an
anti-symmetric tensor operator, $J^{\mu \nu}$,
\beq
P^{\mu}  = (H,\mathbf{P})
\qquad
J^{\mu \nu} =
\left (
\begin{array}{cccc}
0 & -K^1 & -K^2 & -K^3 \\
K^1 & 0 & J^3 & -J^2 \\
K^2 & -J^3 & 0 & J^1 \\
K^3 & J^2 & -J_1 & 0\\ 
\end{array}
\right ) .
\label{c.19}
\eeq
There are two independent polynomial invariants
\beq
M^2 = (P^0)^2 -\mathbf{P}^2 = - P^{\mu}P_{\mu} 
\label{c.20}
\eeq
and
\beq
W^2 = W^{\mu}W_{\mu}
\qquad
W^{\mu} = {1 \over 2} \epsilon^{\mu \nu \alpha \beta} P_{\nu}J_{\alpha \beta}
\label{c.21}
\eeq
where $W^{\mu}$ is the Pauli-Lubanski vector.
When $M\not=0$ the square of the spin is defined by
\beq
S^2 = W^2/M^2 .
\label{c.22}
\eeq
A spin vector $\mathbf{s}$ can be defined by an {\it operator}
rotationless (canonical) 
boost that
transforms the angular momentum tensor to the rest frame:
\beq
s^i = {1 \over 2} \sum_{j,k} \epsilon_{ijk} 
\Lambda_c^{-1} (P)^{j}{}_{\mu}
\Lambda_c^{-1} (P)^{k}{}_{\nu}J^{\mu \nu} 
\label{c.23}
\eeq
where
\beq
\Lambda_c (P)^{\mu}{}_{\nu} =
\left (
\begin{array}{cc}
V^0 & \mathbf{V} \\
\mathbf{V} & I + {\mathbf{V}\otimes \mathbf{V} \over 1+ V^0} \\
\end{array}  
\right )
\qquad V^{\mu}= P^{\mu}/M = {1 \over 2}\mbox{Tr} (B_c (\pmb{\rho})
\sigma_{\mu} B_c (\pmb{\rho}) \sigma_{\nu})
\label{c.24}
\eeq
and $P^{\mu}$, $M$ and $\pmb{\rho}$  are
considered operators related by
\beq
\mathbf{V} = \mathbf{P}/M =  \hat{\bm{\rho}} \sinh (\rho).
\eeq
The spin vector defined with the rotationless boost is called the canonical spin; other types
of spin vectors (helicity, light-front spin) are
related to the canonical spin by
momentum-dependent rotations.  For the purpose of this work it is sufficient to
consider the canonical spin.
The canonical spin can also be expressed in terms of the Pauli-Lubanski vector:
\beq
\left
(  
\begin{array}{c}
0 \\
\mathbf{s}_c \\
\end{array}
\right ) =
 {1 \over M } \Lambda_c^{-1} (P)^{\mu}{}_{\nu} W^{\mu}. 
\label{c.25}
\eeq
where again $\Lambda_c^{-1} (P)^{\mu}{}_{\nu}$ is a matrix of operators.
The components of the spin satisfy $SU(2)$ commutation relations:
\beq
[s_i,s_j] = i \epsilon_{ijk} s^k .
\label{c.27}
\eeq
With these definitions, for $M>0$, $M^2,s^2,\mathbf{P},s_z$ are a
maximal set of commuting self-adjoint functions of the Poincar\'e
generators.  The spectrum of each component of $\mathbf{P}$ is the
real line since each component of $\mathbf{P}$ can be boosted to any
value.  Similarly the spectrum of spins are restricted to be integral
or half integral as a consequence of the $SU(2)$ commutation
relations (\ref{c.27}).  In a general system these commuting observables are not
complete; they can be supplemented by additional Poincar\'e-invariant
degeneracy quantum numbers, which will be denoted by $\alpha$.  A
basis for the Hilbert space are the simultaneous eigenstates of
$M,S^2,\alpha,\mathbf{P},s_z $:
\beq
\{ \vert (m,s,\alpha)
\mathbf{p},\mu \rangle
\}.
\label{c.28}
\eeq
Because these vectors are constructed out of eigenvalues of functions of
$P^{\mu}$ and $J^{\mu \nu}$, which have well-defined Poincar\'e
transformation properties,  the Poincar\'e transformation properties
of these basis states follow from the definitions
\beq
U(A,y)\vert(m,s,\alpha) p,\mu \rangle =e^{i\Lambda p\cdot y}\vert
(m,s,\alpha)\Lambda p,\nu
\rangle
D_{\nu\mu}^{j}[R_{cw}(\Lambda,p)] \sqrt{\frac{\omega_{m}(\Lambda p)}{\omega_{m}(p)}}
\label{c.29}
\eeq
where, $R_{cw}(\Lambda,p):=B_c^{-1}(\Lambda p)\Lambda B_c(p)$ is the
canonical-spin Wigner rotation, $B_c(p)= e^{{1\over 2}\bm{\rho}\cdot \bm{\sigma}}$
where $\bm{\rho}$ is the rapidity of a particle of mass $m$ and momentum
$\mathbf{p}$, 
and $\omega_{m}(p):=\sqrt{m^2+\mathbf{p}^2}$ is the energy of the system.
The square root factors ensure that $U(\Lambda,a)$ is unitary for
states (\ref{c.28}) with the normalization
\beq
\langle (m',s')p',\mu'
\vert (m,s) p,\mu \rangle = \delta_{m'm}\delta_{s's} \delta
(\mathbf{p}'-\mathbf{p}) \delta_{\mu' \mu}.
\label{d.1}
\eeq
The Wigner $D$-function is the finite dimensional unitary representation of the
rotation group in the $\vert s,\mu\rangle $ basis \cite{rose}:
\[
D_{\mu,\mu'}^{s}[R]=\langle s,\mu\vert U(R) \vert s,\mu'\rangle  =
\]
\[
\sum_{k=0}^{s+\mu}\frac{\sqrt{(s+\mu)!(s+\mu')!(s-\mu)!(s-\mu')!}}{k!
  (s+\mu'-k)!(s+\mu-k)!(k-\mu-\mu')!}R_{++}^{k}R_{+-}^{s+\mu'-k}
R_{-+}^{s+\mu-k}R_{--}^{k-\mu-\mu'}
\]
where
\beq
R=
\left (
\begin{array}{cc}
R_{++} & R_{+-}\\
R_{-+} & R_{--}
\end{array}
\right ) = e^{{i\over 2}\bm{\theta}\cdot \bm{\sigma}}=
\sigma_0 \cos ({\theta \over 2})  + i \hat{\bm{\theta}}\cdot \bm{\sigma}
\sin ({\theta \over 2})
\label{c.30}
\eeq
is a $SU(2)$ matrix.  Because $D^s_{\mu \nu}[R]$ is a degree $2s$
polynomial in the matrix elements of $R$, and
$R=e^{i {\bm{\theta}\cdot \bm{\sigma}\over 2}}$ is an entire function of the
angles, $\bm{\theta}$,  it follows that
$D_{\mu,\mu'}^{s}[e^{i {\bm{\theta}\cdot \bm{\sigma}\over 2}}]$ is
an entire function of all three components of $\bm{\theta}$.  This means
that the group representation property
\beq
\sum_{\mu''} D_{\mu,\mu''}^{s}[R_2] D_{\mu'',\mu'}^{s}[R_1] -
D_{\mu,\mu'}^{s}[R_2R_1]=0, 
\label{c.31}
\eeq
and the formulas for adding angular momenta
\beq
D_{\mu,\mu'}^{s}[R] -
\sum_{ \mu_1 \mu_2  \mu_1' \mu_2'}
\langle s,\mu\vert s_1, \mu_1, s_2, \mu_2 \rangle 
D_{\mu_1,\mu_1'}^{s_1}[R] D_{\mu_2,\mu_2'}^{s_2}[R]
\langle s_1, \mu_1', s_2', \mu_2' \vert s , \mu'\rangle =0
\label{c.32}
\eeq
and
\beq
D_{\mu_1,\mu_1'}^{s_1}[R] D_{\mu_2,\mu_2'}^{s_2}[R] -
\sum_{s \mu \mu'} \langle s_1, \mu_1, s_2, \mu_2 \vert s , \mu\rangle
D_{\mu,\mu'}^{s}[R] 
\langle s,\mu'\vert s_1, \mu_1', s_2, \mu_2' \rangle =0,
\label{c.33}
\eeq
which hold for real angles, can be analytically continued to complex
angles.  This means that (\ref{c.31}-\ref{c.33}) also hold when the
$SU(2)$ matrices $R$ are
replaced by $SL(2,\mathbb{C})$ matrices $A$.
In these expressions, $\langle s,\mu\vert s_1, \mu_1, s_2,
\mu_2 \rangle$, are $SU(2)$ Clebsch-Gordan coefficients.  While the
analytic continuation preserves the group representation (\ref{c.31})
and angular
momentum addition (\ref{c.32}-\ref{c.33}) properties, it does not
preserve unitarity.

\section{Euclidean Representations}

In this section the Poincar\'e irreducible basis states (\ref{c.28}) are use
to construct equivalent Euclidean representations of the
irreducible representations of the Poincar\'e group.

The starting point is the irreducible representations of the Poincar\'e
group constructed in the previous section.
The basis vectors and action of $U(\Lambda,a)$ on the basis vectors
are given by (\ref{c.28}) and (\ref{c.29}).

Because $R^{-1} = R^{\dagger}$ for $R \in SU(2)$ the $SU(2)$
representation of the Wigner rotation, $R_{wc}(\Lambda,p)$, can be
expressed in two equivalent ways:
\beq
R_{wc}(\Lambda,p) =
B_c^{-1} (\Lambda p) A B_c (p) = B_c^{\dagger}
(\Lambda p) (A^{\dagger})^{-1} B_c^{\dagger -1} (p) .
\label{e.1}
\eeq
The $SL(2,\mathbb{C})$ group representation property (\ref{c.31})
implies that
the unitary representation $D^s_{\nu \mu} [R_{wc}(\Lambda,p)]$
of the Wigner rotation $R_{wc}(\Lambda,p)$ can be
factored in two different ways:
\beq
D_{\nu\mu}^{s}[R_{wc}(\Lambda,p)]=
\sum_{\alpha \beta} D_{\nu\alpha}^{s}[B_c^{-1} (\Lambda p)]
D_{\alpha\beta}^{s}[A] D_{\beta\mu}^{s}[B_c (p)]
\label{e.2}
\eeq
or
\beq
D_{\nu\mu}^{s}[R_{wc}(\Lambda,p)]= \sum_{\alpha \beta}
D_{\nu\alpha}^{s}[B_c^{\dagger} (\Lambda p)]
D_{\alpha\beta}^{s}[(A^{\dagger})^{-1}]
D_{\beta\mu}^{s}[(B_c^{\dagger})^{-1} (p)] .
\label{e.3}
\eeq
These relations and the group representation properties
(\ref{c.31}- \ref{c.33}) can be used to express equation (\ref{c.29}) in
terms of new Lorentz covariant basis states:
\[
U(A,y) \underbrace{ \sum_\alpha
\vert (m,j) p,\alpha \rangle
D_{\alpha\mu}^{s}[B_c^{-1} (p)] \sqrt{\omega_m(p)}}_
{\vert (m,j) p,\mu \rangle_{cov}}
=
\]
\beq
e^{i\Lambda p\cdot y}\sum_\beta
\underbrace{\sum_\alpha \vert (m,j)\Lambda p,\alpha \rangle 
D_{\alpha\beta}^{s}[B_c^{-1} (\Lambda p)]
\sqrt{\omega_{m}(\Lambda p)}}_
{\vert (m,j) \Lambda p,\beta \rangle_{cov}}
D_{\beta\mu}^{s}[A]
\label{e.4}
\eeq
or 
\[
U(A,y) \underbrace{\sum_\alpha
\vert (m,j) p,\alpha \rangle
D_{\alpha\mu}^{s}[B_c^{\dagger} (p)] \sqrt{\omega_m(p)}}_
{\vert (m,j) p,\mu \rangle_{cov*}}
=
\]
\beq
e^{i\Lambda p\cdot y}\sum_\beta
\underbrace{\sum_\alpha \vert (m,j)\Lambda p,\alpha \rangle
D_{\alpha\beta}^{s}[B_c^{\dagger} (\Lambda p)]
\sqrt{\omega_{m}(\Lambda p)}}_
{\vert (m,j) \Lambda p,\beta \rangle_{cov*}}
D_{\beta\mu}^{s}[(A^{\dagger})^{-1}].
\label{e.5}
\eeq
These expressions replace the states (\ref{c.28}) that transform
covariantly with respect to the Poincar\'e group with states that
transform covariantly with respect to $SL(2,\mathbb{C})$:
\beq
U(A,y) \vert (m,j) p,\mu \rangle_{cov} =
e^{i \Lambda p \cdot y}\sum_{\nu} \vert (m,j)\Lambda  p,\nu \rangle_{cov} D_{\nu\mu}^{s}[A]
\label{e.6}
\eeq
\beq
U(A,y) \vert (m,j) p,\mu \rangle_{cov*} =
e^{i \Lambda p \cdot y}\sum_{\nu} \vert (m,j)\Lambda  p,\nu \rangle_{cov*}
D_{\nu\mu}^{s}[(A^{\dagger})^{-1}]
\label{e.7}
\eeq
These will be referred to as Lorentz covariant representations while
the representations (\ref{c.29}) will be referred to as Poincar\'e
covariant representations.  The transformations relating the Lorentz
and Poincar\'e covariant representations are invertible,
\beq
\vert (m,j)p,\mu \rangle =
\sum \vert (m,j) p,\nu \rangle_{cov}
{1 \over \sqrt{\omega_m (p)}} D^s_{\nu \mu}
[B_c(p)] 
\label{e.8}
\eeq
\beq
\vert (m,j) p,\mu \rangle =
\sum \vert (m,j) p,\nu \rangle_{cov *}
{1 \over \sqrt{\omega_m (p)}} D^s_{\nu \mu}
[(B_c^{\dagger})^{-1}(p)] 
\label{e.9}
\eeq
however there are two distinct Lorentz covariant
representations, because while $R=(R^{\dagger})^{-1}$ for $R\in
SU(2)$, the corresponding representations in $SL(2,\mathbb{C})$ are
inequivalent.  These two representations are called right and
left-handed representations for reasons that will become apparent.

In the Lorentz covariant representations, (\ref{e.4}) and (\ref{e.5}),
this equivalence can be used to show that the equivalent Hilbert space
inner product of two $SL(2,\mathbb{C})$ covariant wave functions has a
non-trivial kernel
\[
\langle \psi \vert \phi \rangle  =
\sum_\mu \int
\langle \psi \vert (m,j) p,\mu \rangle
d\mathbf{p} \langle (m,j) p,\mu \vert \phi \rangle=
\]
\beq
\int \sum_{\mu \nu} \langle \psi \vert (m,j) p,\mu \rangle_{cov}{}
D_{\mu\nu}^{j}[p\cdot\sigma]
2 \delta (p^2+m^2)\theta (p^0) d^4 p{}
_{cov}\langle (m,j) p,\nu \vert \phi \rangle
\label{e.10}
\eeq
\[  
\langle \psi \vert \phi \rangle  =
\int \sum_\mu
\langle \psi \vert (m,j) p,\mu \rangle
d\mathbf{p} \langle (m,j) p,\mu \vert \phi \rangle=
\]
\beq
\int \sum_{\mu \nu}  \langle \psi \vert (m,j) p,\mu \rangle_{cov*}{}
D_{\mu\nu}^{j}[\Pi p\cdot\sigma]
2 \delta (p^2+m^2)\theta (p^0) d^4p{}
_{cov*}\langle (m,j) p,\nu \vert \phi \rangle
\label{e.11}
\eeq
where $B_c (p) B^{\dagger}_c (p)=B_c (p)^2 = {\sigma}\cdot {p}$ and
$B^{-1}_c (p) (B^{\dagger}_c)^{-1}(p) =B^{-2}_c (p) = \Pi {p} \cdot
{\sigma}$, was used in these equations.  $\Pi$ is the space reflection
operator and $p\cdot\sigma = \omega_m(p)\sigma_0 + \mathbf{p}\cdot
\bm{\sigma}$.  These equations explain why (\ref{e.10}) and (\ref{e.11})
are called right and left handed representations.  These kernels are,
up to normalization and change of representation, spin-$s$ two-point
Wightman functions \cite{Wightman:1980} (see eq.  1.55-1.57).

While both the left and right handed representations are each related to
the original Poincar\'e covariant representation, the kernels of the
Lorentz covariant representations of the Hilbert space inner product
do not commute with space reflection.  Instead the right(left)
handed kernel gets mapped into the left(right) handed kernel under
space reflection.

More general classes of spinor representation can be constructed
using tensor products 
\beq
\vert (m,j;j_1,j_2) p,\mu_1,\mu_2  \rangle_{cov} :=
\sum \vert (m,j) p,\nu \rangle
\langle j, \nu \vert j_1, \nu_1, j_2 ,\nu_2 \rangle
\sqrt{\omega_m (p)} 
D^{j_1}_{\nu_1 \mu_1} [ B_c(p)^{-1}]
D^{j_2}_{\nu_2 \mu_2} [ B_c(p)^{\dagger}]
\label{e.12}
\eeq
or direct sums of right and left handed
representations
\beq
\langle (m,j ) p,\nu_1,\nu_2 \vert \phi \rangle_{cov} :=
{1 \over \sqrt{2}}
\langle (m,j) p,\nu \vert \phi \rangle
\sqrt{\omega_m (p)}
\left (
\begin{array}{c}  
D^{j}_{\nu \mu_1} [ B_c(p)^{-1}]\\
D^{j}_{\nu \mu_2} [ B_c(p)^{\dagger}]
\end{array}
\right ).
\label{e.13}
\eeq
Dirac spinors are direct sums of $j=1/2$ left and right handed spinors while
4 vectors are tensor products of $j=1/2$ left and right handed spinors.
The discussion that follows considers the right and left handed
representations separately.  General covariant representations can
be built from the right and left handed representations.
  
The motivation for considering these $SL(2,\mathbb{C})$ covariant
representations is that they are directly related to the
corresponding Euclidean covariant representations.  

A dense set of Hilbert space vectors in the Euclidean representation
are represented by Schwartz functions, $f(x_e,\mu)$ and $g(y_e,\nu)$,
of Euclidean space-time variables, $x_e$ and $y_e$
with positive Euclidean-time support
and spins.   In general the spins are assumed to transform under
finite dimensional representations of $SU(2) \times SU(2)$.

The Euclidean time reflection operator, $\theta$, is defined by
\beq
f(\theta x_e,\mu) = f(\theta (x_e^0,\mathbf{x}),\mu ):= 
f ((-x_e^0,\mathbf{x}),\mu ).
\label{e.14}
\eeq

Consider the following Euclidean covariant kernel:
\beq
S^s_{e}(x_e,\mu;y_e,\nu):=
\int d^4p
{2 \over (2 \pi)^4} {e^{i p_e \cdot (x_e -y_e)}
\over p_e^2 + m^2}D^s_{\mu \nu}(p_e \cdot \sigma_e) .
\label{e.15}
\eeq
This is a distribution since the integral is
not convergent, however it makes perfect sense when considered as the
kernel of a quadratic form.

The physical Hilbert space inner product for a particle of mass $m$ and spin
$s$ is defined by the sesquilinear form
\[
\int\sum_{\mu \nu} d^4x_ed^4y_e f^*(x_e,\mu)
S^s_{e}(\theta x_e,\mu;y_e,\nu)g(y_e,\nu) =
\]
\[
\int\sum_{\mu \nu} d^4p_e f^*(\theta x_e,\mu)
{2 \over (2 \pi)^4} {e^{i p_e \cdot (x_e -y_e)}
\over p_e^2 + m^2}D^s_{\mu \nu}(p_e \cdot \sigma_e)
g(y_e,\nu) =
\]
\beq
\int \sum_{\mu\nu} \psi_{cov}^* (\mathbf{p},\mu) {d \mathbf{p} \over \omega_m (\mathbf{p})} 
D^s_{\mu \nu}(p \cdot \sigma) \phi_{cov}(\mathbf{p},\nu)
\label{e.16}
\eeq
where 
\beq
\psi^*_{cov}(\mathbf{p},\mu):=   
{1 \over (2 \pi)^{3/2}} \int d\mathbf{x}d\tau e^{i \mathbf{p} \cdot \mathbf{x} - \omega_m (\mathbf{p}) x_e^0} f^*(\mathbf{x},x_e^0,\mu) 
\label{e.17}
\eeq
and 
\beq
\phi_{cov}(\mathbf{p},\nu):= 
{1 \over (2 \pi)^{3/2}} \int d^4x_e  e^{-i \mathbf{p} \cdot \mathbf{x} - \omega_m (\mathbf{p}) x_e^0} g(\mathbf{x},x_e^0,\nu).
\label{e.18}
\eeq
The Euclidean time-support condition ensures that the Laplace
transforms with respect to the Euclidean times in (\ref{e.17}) and
(\ref{e.18}) are well defined.  The resulting kernel in (\ref{e.16})
is identical to the Lorentz covariant kernel in (\ref{e.10}) after performing
the integrals over the $p^0_e$.  The covariant wave functions
(\ref{e.17}) and (\ref{e.18}) are
related to the Poincar\'e covariant wave functions by
\beq
\phi(\mathbf{p},\mu) =
D^s_{\mu\nu} (B_c(p))
\phi_{cov}(\mathbf{p},\nu) {1 \over \sqrt{\omega_m(\mathbf{p})}}
\label{e.19}
\eeq
and
\beq
\psi^* (\mathbf{p},\mu) = 
\psi_{cov}^* (\mathbf{p},\nu) D^s_{\nu\mu} (B_c(p))
{1 \over \sqrt{\omega_m(\mathbf{p})}}.
\label{e.20}
\eeq
This shows that the ``Euclidean'' inner product (\ref{e.16}) can be
identified with the corresponding Lorentz covariant inner product,
which itself is identical to the original Poincar\'e covariant inner
product.  These steps illustrate how the correct Minkowski inner
product is obtained from the Euclidean expression without analytic
continuation.

This means that 
\beq
S^s_r(x_e, \mu ; y_e, \nu):= \int
{2 d^4p \over (2 \pi)^4} {e^{i p_e \cdot (x_e -y_e)}
\over p_e^2 + m^2}D^s_{\mu \nu}(p_e \cdot \sigma_e)
\label{e.21}
\eeq
is a Euclidean covariant reflection positive kernel for right
handed representations of mass $m$ and spin $s$ respectively.  The corresponding
kernel for left-handed representations is 
\beq
S^s_l(x_e, \mu ; y_e, \nu):= \int 
{2 d^4 p\over (2 \pi)^4} {e^{i p_e \cdot (x_e -y_e)}
\over p_e^2 + m^2}D^s_{\mu \nu}(\Pi p_e \cdot \sigma_e).
\label{e.22}
\eeq
Space reflection interchanges right and left-handed representations.
The space reflection operator does not commute with the Euclidean
covariant kernel.  This implies that space reflected states will not
transform correctly under Lorentz transformations in these Lorentz
covariant representations.  Kernels for systems that allow a linear
representation of space reflection can be constructed by taking direct
sums or tensor products of right and left handed kernels (see \ref{e.12}-\ref{e.13}). 

The kernels (\ref{e.21}-\ref{e.22}) can be evaluated analytically
using the methods in
\cite{bogoliubov}.
The results are
\[
S^s_{r} (z_e, \mu , \nu) := 
{2 \over (2\pi)^4} \int {d^4p \over p_e^2 + m^2}
D^s_{\mu \nu}(p_e \cdot \sigma_e)
e^{i p_e \cdot z_e} =
\]
\beq
D^s_{\mu \nu}(-i \nabla_{ze} \cdot \sigma_e)
{2m^2
\over (2 \pi)^2}{K_1 (m\sqrt{z_0^2 +\mathbf{z}^2})
\over m\sqrt{z_0^2 +\mathbf{z}^2}}
\label{e.23}
\eeq
\[
S^s_{l} (z_e ,\mu , \nu) := 
{2 \over (2\pi)^2} \int {d^4p  \over p_e^2 + m^2}
D^s_{\mu \nu}(\Pi p_e \cdot \sigma_e)e^{i p_e \cdot z_e} =
\]
\beq
D^s_{\mu \nu}(-i \Pi  \nabla_{ze} \cdot \sigma_e)
{2m^2
\over (2 \pi)^2}{K_1 (m\sqrt{z_0^2 +\mathbf{z}^2})
\over m\sqrt{z_0^2 +\mathbf{z}^2}}
\label{e.25}
\eeq
where $z_e = x_e-y_e$ and $K_1(x)$ is a modified Bessel function. 
Note that ${K_1 (\eta)
\over \eta}$ behaves like $1/\eta^2$ near the origin.   
Since 
$D^s_{\mu \nu}(-i \nabla_{ze} \cdot \sigma_e )$
is a degree $2s$ polynomial in $-i \nabla_{ze}$, these kernels
have power law singularities at the origin, but fall off exponentially
for large values of $z_e^2$. The restriction of the support of the
vectors to positive Euclidean time ensures that $z_e^2 >0$, so the
singularities at $z_e=0$ never cause a problem.  These kernels 
are reflection positive on this space.
This is because $D^s_{\mu \nu}(p\cdot \sigma)
$ factors into a product of a matrix and its adjoint:
\beq
D^s_{\mu \nu}(p\cdot \sigma)
= \sum_{\alpha}D^s_{\mu\alpha}(B_c(p))D^s_{\alpha\nu}(B_c(p))^{\dagger} .
\label{e.26}
\eeq
For any given spin the
$D^s_{\mu \nu}(-i \nabla_{ze} \cdot \sigma_e)$ can be expressed in terms of
higher order modified Bessel functions.

The Euclidean inner product in right and left handed
representations can be expressed directly in the
$x$ representation:
\beq
\langle f \vert g \rangle =
\sum_{\mu \nu} \int f^* ( x,\mu )D^s_{\mu \nu}(-i \nabla_{x} \cdot \sigma_e)
{2m^2
\over (2 \pi)^2}{K_1 (m\sqrt{(\theta x-y)_e^2 }
\over m\sqrt{(\theta x-y)_e^2 }}g(y,\nu) d^4x d^4y 
\label{e.27}
\eeq
\beq
\langle f \vert g \rangle =
\sum_{\mu \nu} \int f^* ( x,\mu )D^s_{\mu \nu}(-i \Pi  \nabla_{x} \cdot \sigma_e)
{2m^2
\over (2 \pi)^2}{K_1 (m\sqrt{(\theta x-y)_e^2 }
\over m\sqrt{(\theta x-y)_e^2 }}g(y,\nu) d^4x d^4y .
\label{e.28}
\eeq
The construction in this section demonstrated the equivalence of the
Poincar\'e covariant, Lorentz covariant and Euclidean covariant
Hilbert space inner products for massive particles with any spin.
Analytic continuation is not used to compute the physical inner
product in the Euclidean representation.  In addition, the Euclidean
inner product, with the Euclidean time reflection on the final state,
and the projection on the space of functions with positive time
support was shown to be non-negative, which demonstrates that these Euclidean
kernels are reflection positive for any spin.

\section{Relativistic invariance - particles}

The formulation of relativistic covariance in the Euclidean
representation is a consequence of the relation between the four
dimensional Euclidean group and the associated complex subgroup of the
Lorentz group discussed in section 2.


This relation is used to relate the infinitesimal generators of
Euclidean transformations to the corresponding Poincar\'e generators and
then show that the resulting Poincar\'e generators are self-adjoint
on the physical Hilbert space.  This is not a new result, but it is
desirable to construct explicit representations for the Poincar\'e
generators for any spin in order to understand the relativistic transformation
properties of particles with different spins or projections of multi-particle
states on irreducible subspaces.

The starting point is to consider the $2 \times 2$ matrix representations
of Minkowski and Euclidean four vectors:
\beq
p \cdot \sigma :=
\left (
\begin{array}{cc}
p^0+p^2 & p^1 -i p^2\\
p^1+i p^2 & p^0 - p^3\\
\end{array}  
\right)  
\qquad 
p_e \cdot \sigma_e :=
\left (
\begin{array}{cc}
i p_e^0+p_e^2 & p_e^1 -i p_e^2\\
p_e^1+i p_e^2 & i p_e^0 - p_e^3\\
\end{array}  
\right).  
\label{p.1}
\eeq
The $SL(2,\mathbb{C})\times SL(2,\mathbb{C})$ transformation properties of
these matrices (denoted by $P$) are
\beq
P\to P' = A P C^t .
\label{p.2}
\eeq
The associated complex $4 \times 4 $ Lorentz and four-dimensional orthogonal
transformation
matrices are
\beq
\Lambda (A,C)^{\mu}{}_{\nu} = {1 \over 2} \mbox{Tr}(\sigma_{\mu}A\sigma_{\nu}C^t)
\qquad
{\cal O} (A,B)^{\mu}{}_{\nu} = {1 \over 2} \mbox{Tr}(\sigma_{e\mu}^{\dagger}A\sigma_{e\nu}C^t) .
\label{p.3}
\eeq
For ordinary rotations
$A=C^*=e^{i{\lambda \over 2} \hat{\mathbf{n}}}$.
For rotations about the $\hat{\mathbf{z}}$ axis 
\beq
{\cal O}(A,A^*)( \lambda) = 
\left (
\begin{array}{cccc}
1 & 0 & 0 & 0 \\
0 & \cos(\lambda)&\sin (\lambda)&0\\
0 & -\sin(\lambda)&\cos (\lambda)&0\\
0 & 0 & 0& 1\\
\end{array}  
\right ).
\label{p.4}
\eeq
These transformations commute with the Euclidean time reflection operator:
\beq
\Theta {\cal O}(A,A^*)( \lambda) \Theta = {\cal O}(A,A^*)( \lambda).
\label{p.5}
\eeq
For real rotations in Euclidean space-time planes, $A=C^t=e^{i{\lambda \over 2} \hat{\mathbf{n}}\cdot \bm{\sigma}}$.
For the case of the $x^0_e-\hat{\mathbf{z}}$ plane
\beq
{\cal O}(A,A^t)( \lambda) x= 
\left (
\begin{array}{cccc}
\cos(\lambda) & 0 & 0 & \sin (\lambda) \\
0 & 1 & 0 &0\\
0 & 0 & 1 &0\\
-\sin (\lambda) & 0 & 0& \cos(\lambda)\\
\end{array}  
\right )
\label{p.6}
\eeq
\and
\beq
\Theta {\cal O}^t(A,A^t)( \lambda) \Theta = {\cal O}(A,A^t)( \lambda).
\label{p.7}
\eeq
While ordinary 3-dimensional rotations are the same for
$p\cdot \sigma$ or $p_e \cdot \sigma_e$,  real rotations in
Euclidean space time planes become
rotationless Lorentz boosts 
with imaginary rapidity when applied to the Minkowski $X$.        

These identifications imply the following algebraic relations between the
infinitesimal generators of the four dimensional orthogonal group and  
the Lorentz group:
\beq
\mathbf{P}_m = \mathbf{P}_e \qquad J^{ij}_{m} = J^{ij}_{e}
\label{p.8}
\eeq
\beq
H_m = i H_e \qquad  K^{i}_{m} = -i J^{0i}_{e}
\label{p.9}
\eeq
Because of the factor of $i$, if the Euclidean generators are
self-adjoint operators on a representation of the Hilbert space, the
Poincar\'e generators (\ref{p.8}) and (\ref{p.9})
cannot be self-adjoint on that representation of the Hilbert space.

In the spinless case ($s=0$) the identifications (\ref{p.3}-\ref{p.7})
result in the following expressions for the infinitesimal generators
of the Poincar\'e group on the Euclidean representation of the
Hilbert space with the Euclidean time reflection:
\beq
H_m\Psi (x_e)= {\partial \over \partial x^0_e}  \Psi (x_e)
\qquad
\mathbf{P}_m\Psi (x_e)= -i {\partial \over \partial \mathbf{x}_e }  \Psi (x_e)
\label{p.10}
\eeq
\beq
\mathbf{J}_m\Psi (x_e)= -i \mathbf{x}
\times \bm{\nabla}_x  \Psi (x_e)
\qquad
K^j_m \Psi (x_e)=( x^j {\partial \over \partial x^0_e}-
x^0_e {\partial \over \partial x^j}) \Psi (x_e).
\label{p.11}
\eeq
It is straightforward to demonstrate that these operators satisfy
the Poincar\'e
commutations relations (\ref{c.16}-\ref{c.18}).  For example
\beq
[K^i_m,H_m] = [ 
x^i {\partial \over \partial x_e^0} -
x^0_e {\partial \over \partial x^i},
{\partial \over \partial x^0_e}]=
i (-i {\partial \over \partial x^i}) =
i P^i_m
\label{p.12}
\eeq
which agrees with (\ref{c.18}). The other commutators can be checked similarly.

The Euclidean time reflection of the final state makes both the
Hamiltonian $H_m$ and the boost generators $\mathbf{K}_m$ formally
Hermitian with respect to the scalar product (\ref{e.16}).  One potential
concern is that even an infinitesimal rotation in a
Euclidean space-time plane can map functions with positive Euclidean
time support to functions that violate the support condition.  This
maps Hilbert space vectors out of the Hilbert space.  The resolution
of this problem will be discussed in section 6.  The subscript $m$ will be
suppressed in what follows.

To show the hermiticity of the rotationless boost generators
(\ref{p.11})
note that rotational invariance of the Euclidean Green's function in
Euclidean space-time planes means that the Euclidean rotation
generators commute with the Euclidean Green's function:
\beq
(-i x^i {\partial \over \partial x_e^0} +i
x^0_e {\partial \over \partial x^i})
S^0_e(x-y) =
S^0_e(x-y)
(-i y^i {\partial \over \partial y_e^0} +i
y^0_e {\partial \over \partial y^i}).
\label{p.13}
\eeq
Multiplying both sides by $i$ gives
\beq
( x^i {\partial \over \partial x_e^0} -
x^0_e {\partial \over \partial x^i})
S^0_e(x-y) =
S^0_e(x-y)
(y^i {\partial \over \partial y_e^0} -
y^0_e {\partial \over \partial y^i}).
\label{p.14}
\eeq
Next consider the inner product
\[
\langle f \vert K^i \vert g \rangle =
\]
\beq
\int d^4x d^4y f^* (\mathbf{x},-x_e^0 ) S^0_e(x-y)
(y^i {\partial \over \partial y_e^0} -
y^0_e {\partial \over \partial y^i}) g(\mathbf{y},y^0_e).
\label{p.15}
\eeq
Using (\ref{p.14}) in (\ref{p.15}) gives  
\beq
= \int d^4x d^4y f^* (\mathbf{x},-x_e^0 )
(x^i {\partial \over \partial x_e^0} -
x^0_e {\partial \over \partial x^i})
S^0_e(x-y)
g(\mathbf{y},y^0_e).
\label{p.16}
\eeq
Integrating by parts again gives
\beq
= -\int d^4x d^4y
(x^i {\partial \over \partial x_e^0} +
x^0_e {\partial \over \partial x^i})
(\theta f)^* (\mathbf{x},x_e^0 )
S^0_e(x-y)
g(\mathbf{y},y^0_e).
\label{p.17}
\eeq
Finally factoring the time reversal out of $f$ gives
\beq
-(x^i {\partial \over \partial x_e^0} +
x^0_e {\partial \over \partial x^i}) \theta 
f^* (\mathbf{x},x_e^0)=
\theta
\left ((x^i {\partial \over \partial x_e^0} -
x^0_e {\partial \over \partial x^i}) 
f^* (\mathbf{x},x_e^0) \right )
\label{p.18}
\eeq
which when used in (\ref{p.17}) gives
\[
\langle f \vert K^i \vert g \rangle =
\int d^4x d^4y f^* (\mathbf{x},-x_e^0 ) S^0_e(x-y)
(y^i {\partial \over \partial y_e^0} -
y^0_e {\partial \over \partial y^i}) g(\mathbf{y},y^0_e) =
\]
\beq
\int d^4x d^4y
\theta ((x^i {\partial \over \partial x_e^0} -
x^0_e {\partial \over \partial x^i}) f (\mathbf{x},x_e^0 ))^*
S^0_e(x-y)
g(\mathbf{y},y^0_e) =
\langle K^i f \vert  g \rangle. 
\label{p.19}
\eeq
This shows that $K^i$ is a Hermitian operator on this representation
of the Hilbert space.

The other non-trivial operator is the Hamiltonian (\ref{p.10}).
In this case
\[
\langle f \vert H \vert g \rangle =
\int d^4x d^4y f^* (\mathbf{x},-x_e^0 ) S^0_e(x-y)
{\partial \over \partial y_e^0}
g(\mathbf{y},y_e^0) =
\]
\[
-\int d^4x d^4y f^* (\mathbf{x},-x_e^0 )
{\partial \over \partial y_e^0}
S^0_e(x-y)
g(\mathbf{y},y^0_e) =
\int d^4x d^4y f^* (\mathbf{x},-x_e^0 )
{\partial \over \partial x_e^0}
S^0_e(x-y)
g(\mathbf{y},y^0_e) =
\]
\beq
-\int d^4x d^4y
{\partial \over \partial x_e^0}
f^* (\mathbf{x},-x_e^0 )
S^0_e(x-y)
g(\mathbf{y},y^0_e) =
\int d^4x d^4y
{\partial f^* \over \partial x^0}
(\mathbf{x},-x_e^0 ) 
S^0_e(x-y)
g(\mathbf{y},y^0_e) =
\langle Hf \vert g \rangle .
\label{p.20}
\eeq
The Euclidean time reversal does not change the linear or angular
momentum operators.  These methods can be used to demonstrate that all
of the $s=0$ generators (\ref{p.10}-\ref{p.11}) are Hermitian in the
Euclidean representation of the Hilbert space and satisfy the Poincar\'e Lie
algebra.

\section{Spin}

For application in hadronic physics or relativistic many-body physics
it is necessary to consider representations of the Poincar\'e Lie
algebra with higher spins.  In this section explicit formulas for
generators for particles with arbitrary spin are derived, generalizing
the method used in the previous section for scalar particles.  While
these results are not new, explicit formulas are needed for
applications.

In the original Poincar\'e covariant theory the spin is associated
with the observable that is the $\hat{\mathbf{z}}$-component of the
spin that would be measured in the particle's rest frame if it was
transformed to the rest frame with a rotationless Lorentz
transformation.  The spin in the covariant wave function is related to
this spin by multiplying by one of the $SL(2,\mathbb{C})$ matrices,
$D^s_{\mu \nu}(B_c(p)^{-1})$ or $D^s_{\mu
  \nu}(B_c(p)^{\dagger})$.  These transformations lead to
distinct right or left handed spinors.  In discussing spin it is
important to understand that the Poincar\'e covariant spinors and the
Lorentz covariant spinors are related, but have
different transformation properties.  Representations
of the Poincar\'e generators for right and left handed covariant spins
must be considered separately. In addition, for each type of covariant spinor
there are invariant linear functionals that define dual spinors.  The
dual spinors are spinor analogs of covariant and contravariant
vectors.  In conventional treatments
\cite{Wightman:1980}\cite{Wightman} \cite{berestetskii} the
right-handed spinors are denoted by $\xi^a$, left handed spinors are
denoted by $\xi^{\dot{a}}$ and their duals are denoted by $\xi_a$ and
$\xi_{\dot{a}}$ respectively.  In this section we consider each of these
four cases.

The first step is to determine the Euclidean covariance properties of the
Euclidean kernels for right and left handed covariant spinors and
their duals.
Euclidean four vectors can be represented by any of the four matrices:
\beq
p_e \cdot \sigma_e = p_e^\mu\sigma_{e\mu}
\qquad
p_e \cdot ( \sigma_2 \sigma_e \sigma_2 )  =  p_e^\mu\sigma_2 \sigma_{e\mu} \sigma_2
\qquad
p_e \cdot \sigma^t_e  =  p_e^\mu\sigma^t_{e\mu}
\qquad
p_e \cdot ( \sigma_2 \sigma_e^t \sigma_2 )
=  p_e^\mu\sigma_2 \sigma^t_{e\mu} \sigma_2 .
\label{s.1}
\eeq
The determinant of each of these matrices is (-) the square of the
Euclidean length of $p_e$, which is preserved under linear 
transformations of the form
\beq
P' = APC^t
\label{s.2}
\eeq
where $P$ represents any of the matrices in (\ref{s.1}), and $A,C \in
SL(2,\mathbb{C})$.  Real four-dimensional orthogonal transformations
are obtained by restricting $A$ and $C$ to be elements of $SU(2)$.

The $4 \times 4$ orthogonal matrix $\mathbb{O}(A,C)^{\mu}{}_{\nu}$
is related to the pair $(A,C)$ by
\beq
\mathbb{O}(A,C)^{\mu}{}_{\nu}:=\frac{1}{2}Tr(\sigma_{e\mu}^\dagger A\sigma_{e\nu}C^t).
\label{s.3}
\eeq
It follows that
\beq
A p_e^{\mu}\sigma_{e\mu} C^t= \sigma_{e\mu}\mathbb{O}(A,C)^{\mu}{}_{\nu}p_e^{\nu}
 = \sigma_{e\mu} (\mathbb{O}(A,C)p_e)^{\mu}.
\label{s.4}
\eeq
Multiplying (\ref{s.4}) by $\sigma_2$ on both sides using
$\sigma_2 A \sigma_2 = A^*$ for $A\in SU(2)$ gives
\beq
A^*(p_e \cdot \sigma_2\sigma_e\sigma_2)C^\dagger=(\mathbb{O}(A,C)p)_e\cdot\sigma_2\sigma_e\sigma_2 .
\label{s.5}
\eeq
Taking transposes of the $2 \times 2$ matrices (\ref{s.4}) and (\ref{s.5})
give 
\beq
C(p_e\sigma_e^t)A^t= \sigma_e^{t}\cdot(\mathbb{O}(A,C)p_e) 
\label{s.6}
\eeq
and
\beq
C^*(p_e \cdot \sigma_2\sigma_e^{t}\sigma_2)A^\dagger=
\sigma_2\sigma_e^{t}\sigma_2 \cdot
(\mathbb{O}(A,C)p_e) ,
\label{s.7}
\eeq
In all four of these expressions
$A$, $C$ and the orthogonal matrix
$\mathbb{O}(A,C)$ are unchanged.  All four of the
matrices (\ref{s.1}) become positive when $p_e$ is replaced by the
on-shell Minkowski four momentum,
$p^{\mu}_m = (\sqrt{\mathbf{p}^2+m^2},\mathbf{p})$
and $\sigma_e^{\mu}$ is replaced by $\sigma^{\mu}$. 

These identities will be used to derive the covariance properties
of each type of Euclidean kernel.

The  matrices (\ref{s.1}) appear in the Euclidean covariant
kernels for the right and left-handed representations
and their duals.  The spin $s$ Euclidean covariant inner product
kernels for each type of covariant spinor are:
\beq
S_e^s (x_e;\mu,\nu)={2 \over (2\pi)^4} \int {D^s_{\mu\nu} [p_e \cdot \sigma_{e}]\over p_e^2 +m^2}
e^{i p_e \cdot x_e}d^4p_e
\label{s.8}
\eeq
\beq
S_{ed}^s (x_e;\mu,\nu) ={2 \over (2\pi)^4} \int
{D^s_{\mu\nu}  [p_e \cdot (\sigma_2 \sigma_{e} \sigma_2)]\over p_e^2 +m^2}
e^{i p_e \cdot x_e}d^4p_e
\label{s.9}
\eeq
\beq
S_{e*}^s (x_e;\mu,\nu) ={2 \over (2\pi)^4} \int
{D^s_{\mu\nu}  [p_e \cdot \sigma^t_{e}]\over p_e^2 +m^2}
e^{i p_e \cdot x_e}d^4p_e
\label{s.10}
\eeq
\beq
S_{ed*}^s (x_e;\mu,\nu) ={2 \over (2\pi)^4} \int
{D^s_{\mu\nu}  [p_e \cdot (\sigma_2 \sigma^t_{e} \sigma_2)]\over p_e^2 +m^2}
e^{i p_e \cdot x_e}d^4p_e .
\label{s.11}
\eeq
It is possible to construct more general classes of kernels using
products or direct sums of left and right handed representations,
for example
\beq
S_{e:e*}^{s\dot s} (x_e;\mu,\dot\mu,\nu,\dot\nu) =
{2 \over (2\pi)^4} \int
{D^s_{\mu\nu}  [p_e \cdot (\sigma_2 \sigma_{e} \sigma_2)]
 D^{\dot s}_{\dot\mu\dot\nu}  [p_e \cdot \sigma^t_{e}]
\over p_e^2 +m^2}
e^{i p_e \cdot x_e}d^4p_e .
\eeq

The physical Hilbert space
inner product associated with each of these kernels is
\beq
\langle\psi_{e} \vert \phi_{e} \rangle=
\int \sum_{\mu \nu}\psi_e^*(\theta x,\mu)
S_e^s (x_e-y_e;\mu,\nu) \phi_e(y,\nu)d^4x d^4y
\label{s.12}
\eeq
\beq
\langle\psi_{ed} \vert \phi_{ed} \rangle=
\int \sum_{\mu \nu} \psi^*_{ed}(\theta x,\mu)
S_{ed}^s (x_e-y_e;\mu,\nu) \phi_{ed}(y,\nu)d^4x d^4y
\label{s.13}
\eeq
\beq
\langle\psi_{e*} \vert \phi_{e*} \rangle=
\int \sum_{\mu \nu} \psi^*_{e*}(\theta x,\mu)
S_{e*}^s (x_e-y_e;\mu,\nu) \phi_{e*}(y,\nu)d^4x d^4y
\label{s.14}
\eeq
\beq
\langle\psi_{ed*} \vert \phi_{ed*} \rangle=
\int \sum_{\mu \nu} \psi^*_{ed*}(\theta x,\mu)
S_{ed*}^s (x_e-y_e;\mu,\nu) \phi_{ed*}(y,\nu)d^4x d^4y .
\label{s.15}
\eeq
For wave functions with positive Euclidean time support, the $p^0_e$
integral can be evaluated by the residue theorem, closing the contour
in the upper half plane.  This replaces $p^0_e$ by
$i\omega_m(\mathbf{p})$.  The kernels become the two-point
Minkowski Wightman functions \cite{Wightman:1980} for
mass $m$ spin $s$ irreducible
representations of the Lorentz group.  Equations (\ref{s.12}) and
(\ref{s.13}) are dual representations of the right-handed kernel,
while (\ref{s.14}) and (\ref{s.15}) are dual representations of the
left-handed kernel.  $\sigma_2$ behaves like a metric tensor for the
Lorentz covariant spinors, relating the representations
(\ref{s.12}) and (\ref{s.13}) or (\ref{s.14}) and (\ref{s.15}).
Contraction of the two types of right
or left handed spinors are Lorentz invariant.  The results of performing
the
$p^0_e$ integral for each type of kernel are
\beq
\langle \psi_{e} \vert \phi_{e}\rangle= 
\int \sum_{\mu \nu} f^*_m(\mathbf{p},\mu)
{d\mathbf{p}
D_{\mu \nu}^s[p_m\cdot
\sigma] \over \omega_m (\mathbf{p})}
g_m(\mathbf{p},\nu)
\label{s.16}
\eeq
\beq
\langle \psi_{ed} \vert \phi_{ed} \rangle=
\int \sum_{\mu \nu} f^*_m(\mathbf{p},\mu)
{d\mathbf{p}
D_{\mu \nu}^s[p_m\cdot \sigma_2 \sigma
\sigma_2] \over \omega_m (\mathbf{p})}
g_m(\mathbf{p},\nu)
\label{s.17}
\eeq
\beq
\langle \psi_{e*}\vert \phi_{e*}\rangle=
\int \sum_{\mu \nu}
f^*_m(\mathbf{p},\mu)
{d\mathbf{p}
D_{\mu \nu}^s[p_m\cdot
\sigma^*] \over \omega_m (\mathbf{p})}
g_m(\mathbf{p},\nu)
\label{s.18}
\eeq
\beq
\langle \psi_{ed*} \vert \phi_{ed*}\rangle=
\int \sum_{\mu \nu} f^*_m(\mathbf{p},\mu)
{d\mathbf{p}
D_{\mu \nu}^s[p_m\cdot \sigma_2
\sigma^* \sigma_2 ]  
\over \omega_m (\mathbf{p})}
g_m(\mathbf{p},\nu)
\label{s.19}
\eeq
where
\beq
f^*_m(\mathbf{p},\mu):= 
\int {d^4x \over (2 \pi)^{3/2}}
\psi^*(x,\mu)e^{i \mathbf{p}\cdot \mathbf{x} - \omega_m (\mathbf{p})x^0}
\label{s.20}
\eeq
\beq
g_m(\mathbf{p},\nu):=
{d^4y \over (2 \pi)^{3/2}}
\phi(y,\nu)e^{-i \mathbf{p}\cdot \mathbf{y} - \omega_m (\mathbf{p})y^0}
\label{s.21}
\eeq
for each type of spinor wave function. 

Each of the spin matrices, 
$D_{\mu \nu}^s[p_m\cdot \sigma]$,
$D_{\mu \nu}^s[p_m\cdot \sigma_2 \sigma \sigma_2]$,
$D_{\mu \nu}^s[p_m\cdot \sigma^*]$ and 
$D_{\mu \nu}^s[p_m\cdot \sigma_2 \sigma^* \sigma_2 ]$ are   
positive Hermitian matrices,
so the Euclidean Green's functions (\ref{s.8}-\ref{s.11})
are all reflection positive.

The spinor transformation properties
(\ref{s.4}-\ref{s.7}) of the right and left
handed spinors and their duals are used
construct the spinor parts of the Poincar\'e generators
in the Euclidean representation:
\[
\int \sum_{\mu \nu} \psi_e^*(\theta x,\mu) \frac{e^{ip\cdot(x-y)}}{p^2+m^2}D_{\mu \nu}^s[\mathbb{O}p\cdot \sigma_e]\phi_e(y,\nu)d^4x d^4y d^4p=
\]
\beq
\int \sum_{\mu \nu} \psi_e^*(\theta x,\mu) \frac{e^{ip\cdot(x-y)}}{p^2+m^2}D_{\mu \nu}^s[p\cdot A\sigma_e C^t]\phi_e(y,\nu)d^4x d^4y d^4p
\label{s.22}
\eeq
\[
\int \sum_{\mu \nu} \psi_{ed}^*(\theta x,\mu) \frac{e^{ip\cdot(x-y)}}{p^2+m^2}
D_{\mu \nu}^s[\mathbb{O}p\cdot \sigma_2 \sigma_e \sigma_2]
\phi_{ed}(y,\nu)d^4x d^4y d^4p=
\]
\beq
\int \sum_{\mu \nu} \psi_{ed}^*(\theta x,\mu) \frac{e^{ip\cdot(x-y)}}{p^2+m^2}D_{\mu \nu}^s[p\cdot A^*\sigma_2 \sigma_e \sigma_2C^\dagger]\phi_{ed}(y,\nu)d^4x d^4y d^4p
\label{s.23}
\eeq

\[
\int \sum_{\mu \nu} \psi^*_{e*}(\theta x,\mu) \frac{e^{ip\cdot(x-y)}}{p^2+m^2}D_{\mu \nu}^s[\mathbb{O}p\cdot \sigma_e^t]\phi_{e*}(y,\nu)d^4x d^4y d^4p=
\]
\beq
\int \sum_{\mu \nu} \psi^*_{e*}(\theta x,\mu) \frac{e^{ip\cdot(x-y)}}{p^2+m^2}D_{\mu \nu}^s[p\cdot C\sigma_e^tA^t]\phi_{e*}(y,\nu)d^4x d^4y d^4p
\label{s.24}
\eeq
 
\[
\int \sum_{\mu \nu} \psi^*_{ed*}(\theta x,\mu) \frac{e^{ip\cdot(x-y)}}{p^2+m^2}D_{\mu \nu}^s[\mathbb{O}p\cdot \sigma_2 \sigma_e^t \sigma_2 ]\phi_{ed*}(y,\nu)d^4x d^4y d^4p=
\]
\beq
\int \sum_{\mu \nu} \psi_{ed*}^*(\theta x,\mu) \frac{e^{ip\cdot(x-y)}}{p^2+m^2}D_{\mu \nu}^s[p\cdot C^*\sigma_2 \sigma_e^t \sigma_2A^\dagger ]\phi_{ed*}(y,\nu)d^4x d^4y d^4p .
\label{s.25}
\eeq

The next step is to move the transformations in the kernels to the
wave functions.  The Euclidean invariance of the measures and scalar
products, the group representation properties of the Wigner functions,
and re-definitions of the wave functions can be used to show that
(\ref{s.22}-\ref{s.25}) are equivalent to
\[
\int \sum   (D^s_{\mu\alpha}[A^\dagger]^{-1}{\psi}_e(\theta \mathbb{O}^t \theta  x,\alpha))^* \frac{e^{ip\cdot(\theta x-y)}}{p^2+m^2}D_{\mu \nu}^s[p\cdot \sigma_e]
{\phi}_e(y,\nu)d^4x d^4y d^4p
\]
\beq
=\int \sum  {\psi}_e^*( x,\mu) \frac{e^{ip\cdot(\theta x-y)}}{p^2+m^2}D_{\mu \alpha}^j[p\cdot \sigma_e]D^s_{\alpha \nu}[C^t]{\phi}_e(\mathbb{O}y,\nu)d^4x d^4y d^4p
\label{s.26}
\eeq

\[
\int \sum  ( D^s_{\mu \alpha}[A^t]^{-1}{\psi}_{ed}(\theta \mathbb{O}^t \theta   x,\alpha))^* \frac{e^{ip\cdot(\theta x-y)}}{p^2+m^2}D_{\mu \nu}^s[p\cdot \sigma_2 \sigma_e \sigma_2]{\phi}_{ed}(y,\nu)d^4x d^4y d^4p
\]
\beq
=\int \sum  {\psi}^*_{ed}( x,\mu) \frac{e^{ip\cdot(\theta x-y)}}{p^2+m^2}D_{\mu \alpha}^j[p\cdot \sigma_2 \sigma_e \sigma_2]
D^s_{\alpha \nu}[C^\dagger]{\phi}_{ed}(\mathbb{O}y,\nu)d^4x d^4y d^4p
\label{s.27}
\eeq

\[
\int \sum  (D^s_{\mu \alpha}[C^\dagger]^{-1} {\psi}_{e*}(\theta  \mathbb{O}^t \theta x,\alpha))^* \frac{e^{ip\cdot(\theta x-y)}}{p^2+m^2}D_{\mu \nu}^s[p\cdot \sigma_e^t]{\phi}_{e*}(y,\nu)d^4x d^4y d^4p
\]
\beq
=\int \sum  {\psi}_{e*}^*( x,\mu) \frac{e^{ip\cdot(\theta x-y)}}{p^2+m^2}D_{\mu \alpha}^j[p\cdot \sigma_e^t]D^s_{\alpha \nu}[A^t]\Tilde{\phi}_{e*}(\mathbb{O}y,\nu)d^4x d^4y d^4p
\label{s.28}
\eeq

\[
\int \sum  (D^s_{\mu \alpha}[C^t]^{-1} {\psi}_{ed*}(\theta  \mathbb{O}^t\theta  x,\alpha))^* \frac{e^{ip\cdot(\theta x-y)}}{p^2+m^2}D_{\mu \nu}^s[p\cdot \sigma_2 \sigma_e^t \sigma_2 ]{\phi}_{ed*}(y,\nu)d^4x d^4y d^4p
\]
\beq
= \int \sum  {\psi}_{ed*}^*( x,\mu) \frac{e^{ip\cdot(\theta
    x-y)}}{p^2+m^2}D_{\mu \alpha}^s[p\cdot \sigma_2 \sigma_e^t \sigma_2]
D^s_{\alpha \nu}[A^\dagger]{\phi}_{ed*}(\mathbb{O}y,\nu)d^4x d^4y d^4p .
\label{s.29}
\eeq
For ordinary rotations, as well as rotations in space Euclidean time planes,
the $SU(2)$ matrices $A$ and $C$ are related.

To derive expressions for the generators for each type of spinor,
check the hermiticity and verify the commutation relations
the first step is to replace $A$ and $C$ with
the pairs of $SU(2)$ matrices representing 
one-parameter groups for both ordinary rotations about a fixed
axis and rotations in a Euclidean space time plane.

For ordinary rotations about the $\hat{\mathbf{n}}$ axis,
the one-parameter group is 
\beq
A(\lambda ) = C^*(\lambda) = e^{i {\lambda \over 2}\hat{\mathbf{n}}
\cdot \bm{\sigma}}
\label{s.30}
\eeq
and
$(\theta  \mathbb{O}^t(\lambda)\theta) = \mathbb{O}^t(\lambda)$,
while for rotations in Euclidean $\hat{\mathbf{n}}$-$x^0$  space-time planes
the one-parameter group is
\beq
A(\lambda ) = C^t(\lambda) = e^{i {\lambda \over 2}\hat{\mathbf{n}}
\cdot \bm{\sigma}}
\label{s.31}
\eeq
and
$(\theta  \mathbb{O}^t(\lambda)\theta) = \mathbb{O}(\lambda)$.
The $4 \times 4$ orthogonal transformations, $\mathbb{O}(\lambda)$
associated with each
type of transformation are shown explicitly for
rotations about the $\hat{\mathbf{z}}$ axis and for
rotations in the $\hat{\mathbf{z}}$-$x^0$
plane:
For rotations about the $\hat{\mathbf{z}}$ axis 
\beq
\mathbb{O}(A,A^*)( \lambda) = 
\left (
\begin{array}{cccc}
1 & 0 & 0 & 0 \\
0 & \cos(\lambda)&\sin (\lambda)&0\\
0 & -\sin(\lambda)&\cos (\lambda)&0\\
0 & 0 & 0& 1\\
\end{array}  
\right )
\label{s.32}
\eeq
and
\beq
\Theta \mathbb{O}(A,A^*)( \lambda) \Theta = \mathbb{O}(A,A^*)( \lambda).
\label{s.33}
\eeq
For rotations in the $\hat{\mathbf{z}}$-$x^0$ plane
\beq
\mathbb{O}(A,A^t)( \lambda)= 
\left (
\begin{array}{cccc}
\cos(\lambda) & 0 & 0 & \sin (\lambda) \\
0 & 1 & 0 &0\\
0 & 0 & 1 &0\\
-\sin (\lambda) & 0 & 0& \cos(\lambda)\\
\end{array}  
\right )
\label{s.34}
\eeq
and
\beq
\theta \mathbb{O}^t(A,A^t)( \lambda) \theta = \mathbb{O}(A,A^t)( \lambda)
\label{s.35}
\eeq

For the case of ordinary rotations
$A=C^*$ and equations (\ref{s.26}-\ref{s.29}) become
\[
\int \sum  (D^s_{\mu\alpha}[A]{\psi}_e ( \mathbb{O}^t(\lambda)   x,\alpha) )^*  \frac{e^{ip\cdot(\theta x-y)}}{p^2+m^2}D_{\mu \nu}^s[p\cdot \sigma_e]{\phi}_e(y,\nu)d^4x d^4y d^4p
\]
\beq
=\int \sum  {\psi}_e^*( x,\mu) \frac{e^{ip\cdot(\theta x-y)}}{p^2+m^2}D_{\mu \alpha}^j[p\cdot \sigma_e]D^s_{\alpha \nu}[A^{\dagger}]{\phi}_e(\mathbb{O}(\lambda)y,\nu)d^4x d^4y d^4p
\label{s.36}
\eeq

\[
  \int \sum ( D^s_{\mu \alpha}[A^*]{\psi}_{ed}( \mathbb{O}^t(\lambda)   x,\alpha))^*
  \frac{e^{ip\cdot(\theta x-y)}}{p^2+m^2}D_{\mu \nu}^s[p\cdot \sigma_2 \sigma_e \sigma_2]{\phi}_{ed}(y,\nu)d^4x d^4y d^4p
\]
\beq
=\int \sum  {\psi}_{ed}^*( x,\mu) \frac{e^{ip\cdot(\theta x-y)}}{p^2+m^2}D_{\mu \alpha}^j[p\cdot \sigma_2 \sigma_e \sigma_2]
D^s_{\alpha \nu}[A^t] {\phi}_{ed}(\mathbb{O}(\lambda)y,\nu)d^4x d^4y d^4p
\label{s.37}
\eeq

\[
\int \sum  (D^s_{\mu \alpha}[A^*] {\psi}_{e*}( \mathbb{O}^t(\lambda)   x,\alpha))^*
\frac{e^{ip\cdot(\theta x-y)}}{p^2+m^2}D_{\mu \nu}^s[p\cdot \sigma_e^t]{\phi}_{e*}(y,\nu)d^4x d^4y d^4p
\]
\beq
=\int \sum  {\psi}_{e*}^*( x,\mu) \frac{e^{ip\cdot(\theta x-y)}}{p^2+m^2}D_{\mu \alpha}^j[p\cdot \sigma_e^t]D^s_{\alpha \nu}[A^t]{\phi}_{e*}(\mathbb{O}(\lambda)y,\nu)d^4x d^4y d^4p
\label{s.38}
\eeq

\[
\int \sum (D^s_{\mu \alpha}[A] {\psi}_{ed*}( \mathbb{O}^t(\lambda)   x,\alpha))^*
\frac{e^{ip\cdot(\theta x-y)}}{p^2+m^2}D_{\mu \nu}^s[p\cdot \sigma_2
\sigma_e^t \sigma_2 ]{\phi}_{ed*}(y,\nu)d^4x d^4y d^4p
\]
\beq
= \int \sum {\psi}_{ed*}^*( x,\mu) \frac{e^{ip\cdot(\theta
    x-y)}}{p^2+m^2}D_{\mu \alpha}^s[p\cdot \sigma_2 \sigma_e^t \sigma_2
]D^s_{\alpha \nu}[A^\dagger]{\phi}_{ed*}(\mathbb{O}(\lambda)y,\nu)d^4x d^4y d^4p
\label{s.39}
\eeq

For the case of rotations in Euclidean space-time planes
for $A=C^t$ equations (\ref{s.26}-\ref{s.29}) become
\[
\int \sum  (D^s_{\mu\alpha}[A]{\psi}_e(\mathbb{O}(\lambda)  x,\alpha))^* \frac{e^{ip\cdot(\theta x-y)}}{p^2+m^2}D_{\mu \nu}^s[p\cdot \sigma_e]{\phi_e}(y,\nu)d^4x d^4y d^4p
\]
\beq
=\int \sum  {\psi_e}^*( x,\mu) \frac{e^{ip\cdot(\theta x-y)}}{p^2+m^2}D_{\mu \alpha}^j[p\cdot \sigma_e]D^s_{\alpha \nu}[A]
{\phi_e}(\mathbb{O}(\lambda)y,\nu)d^4x d^4y d^4p
\label{s.40}
\eeq

\[
\int \sum ( D^s_{\mu \alpha}[A^*]{\psi}_{ed}(\mathbb{O}(\lambda)  x,\alpha))^* \frac{e^{ip\cdot(\theta x-y)}}{p^2+m^2}D_{\mu \nu}^s[p\cdot \sigma_2 \sigma_e \sigma_2]{\phi}_{ed}(y,\nu)d^4x d^4y d^4p
\]
\beq
=\int \sum {\psi}_{ed}^*( x,\mu) \frac{e^{ip\cdot(\theta x-y)}}{p^2+m^2}D_{\mu \alpha}^j[p\cdot \sigma_2 \sigma_e \sigma_2]
D^s_{\alpha \nu}[A^*]{\phi}_{ed}(\mathbb{O}y,\nu)d^4x d^4y d^4p
\label{s.41}
\eeq

\[
\int \sum (D^s_{\mu \alpha}[A^t] {\psi}_{e*}(\mathbb{O}(\lambda)  x,\alpha))^* \frac{e^{ip\cdot(\theta x-y)}}{p^2+m^2}D_{\mu \nu}^s[p\cdot \sigma_e^t]{\phi}_{e*}(y,\nu)d^4x d^4y d^4p
\]
\beq
=\int \sum {\psi}_{e*}^*( x,\mu) \frac{e^{ip\cdot(\theta x-y)}}{p^2+m^2}D_{\mu \alpha}^j[p\cdot \sigma_e^t]D^s_{\alpha \nu}[A^t]{\phi}_{e*}(\mathbb{O}(\lambda)y,\nu)d^4x d^4y d^4p
\label{s.42}
\eeq

\[
\int \sum (D^s_{\mu \alpha}[A^\dagger ] {\psi}_{ed*}(\mathbb{O}(\lambda)  x,\alpha))^*
\frac{e^{ip\cdot(\theta x-y)}}{p^2+m^2}D_{\mu \nu}^s[p\cdot \sigma_2 \sigma_e^t \sigma_2 ]{\phi}_{ed*}(y,\nu)d^4x d^4y d^4p
\]
\beq
= \int \sum  {\psi}_{ed*}^*( x,\mu) \frac{e^{ip\cdot(\theta
    x-y)}}{p^2+m^2}D_{\mu \alpha}^s[p\cdot \sigma_2 \sigma_e^t \sigma_2
]D^s_{\alpha \nu}[A^\dagger]{\phi}_{ed*}(\mathbb{O}(\lambda)y,\nu)d^4x d^4y d^4p.
\label{s.43}
\eeq

To construct generators of ordinary rotations differentiate the
right hand side of
(\ref{s.36}-\ref{s.39}) by $\lambda$, set $\lambda=0$, and multiply the result
by $i$.   To construct the generators of Euclidean space-time rotations
differentiate the right hand side of
(\ref{s.40}-\ref{s.43}) by $\lambda$, set $\lambda=0$, and multiply the
result by $i$
to get expressions for the generators.  To get expressions for the
Lorentz boost generators multiply the Euclidean space-time rotation
generators by an additional factor of $-i$. 
The derivatives of the
Wigner functions can be computed using
\beq
{d \over d\lambda}D^s_{\mu \nu} [A(\lambda)]_{\vert_{\lambda=0}} =
{d \over d\lambda}\langle s, \mu \vert
e^{i \lambda \hat{\mathbf{n}}\cdot \mathbf{S}}
\vert s, \nu \rangle_{\vert_{\lambda=0}}=
i \langle s, \mu \vert \hat{\mathbf{n}}\cdot \mathbf{S} \vert s, \nu \rangle
\label{s.45}
\eeq
\beq
{d \over d\lambda}D^s_{\mu \nu} [A(\lambda)^{\dagger}]_{\vert_{\lambda=0}} =
{d \over d\lambda}\langle s, \mu \vert e^{-i \lambda \hat{\mathbf{n}}\cdot \mathbf{S}} \vert s, \nu \rangle_{\vert_{\lambda=0}} =
- i \langle s, \mu \vert \hat{\mathbf{n}}\cdot \mathbf{S} \vert s, \nu \rangle
\label{s.46}
\eeq
\beq
{d \over d\lambda}D^s_{\mu \nu} [A^*(\lambda)]_{\vert_{\lambda=0}} =
{d \over d\lambda}(D^s_{\mu \nu} [A(\lambda)])^*_{\vert_{\lambda=0}} =
-i \langle s, \mu \vert \hat{\mathbf{n}}\cdot \mathbf{S} \vert s, \nu \rangle^*=
-i \langle s, \nu \vert \hat{\mathbf{n}}\cdot \mathbf{S} \vert s, \mu \rangle
\label{s.47}
\eeq
\beq
{d \over d\lambda}D^s_{\mu \nu} [A^t(\lambda)]_{\vert_{\lambda=0}} =
{d \over d\lambda}(D^s_{\mu \nu} ([A(\lambda)])^*)^{-1}_{\vert_{\lambda=0}} =
i\langle s, \mu \vert \hat{\mathbf{n}}\cdot \mathbf{S} \vert s, \nu \rangle^*=
i\langle s, \nu \vert \hat{\mathbf{n}}\cdot \mathbf{S} \vert s, \mu \rangle .
\label{s.48}
\eeq
These can be evaluated using $S_z$ and angular momentum
raising and lowering operators.
The rotation generators for each type of spinor representation can be read off
of (\ref{s.36}-\ref{s.39}):
\beq
\langle x,s,\nu \vert  \mathbf{J} \vert \psi_e \rangle =
\sum_\nu \left (\delta_{\mu \nu} \mathbf{x} \times (-i {\partial \over \partial \mathbf{x}})
+ \langle s, \mu \vert \hat{\mathbf{n}}\cdot \mathbf{S} \vert s, \nu \rangle
\right ) \langle x,s,\nu \vert \psi_e \rangle 
\label{s.49}
\eeq
\beq
\langle x,s,\nu \vert  \mathbf{J} \vert \psi_{ed} \rangle =  
\sum_\nu \left ( \delta_{\mu \nu} \mathbf{x} \times (-i {\partial \over \partial \mathbf{x}})
- \langle s, \nu \vert \hat{\mathbf{n}}\cdot \mathbf{S} \vert s, \mu \rangle
\right ) \langle x,s,\nu \vert \psi_{ed} \rangle
\label{s.50}
\eeq
\beq
\langle x,s,\nu \vert  \mathbf{J} \vert \psi_{e*} \rangle =  
\sum_\nu \left ( \delta_{\mu \nu} \mathbf{x} \times (-i {\partial \over \partial \mathbf{x}})
- \langle s, \nu \vert \hat{\mathbf{n}}\cdot \mathbf{S} \vert s, \mu \rangle
\right ) \langle x,s,\nu \vert \psi_{e*} \rangle
\label{s.51}
\eeq
\beq
\langle x,s,\nu \vert  \mathbf{J} \vert \psi_{ed*} \rangle =  
\sum_\nu \left ( \delta_{\mu \nu} \mathbf{x} \times (-i {\partial \over \partial \mathbf{x}})
+ \langle s, \mu \vert \hat{\mathbf{n}}\cdot \mathbf{S} \vert s, \nu \rangle
\right ) \langle x,s,\nu \vert \psi_{ed*} \rangle
\label{s.52}
\eeq
The first and fourth term are representations of standard rotation generators
.  In the second and third terms the spin generator matrix elements are
transposed and multiplied by
with a (-) sign.  To show that these operator satisfy $SU(2)$
commutation relations, 
consider matrices satisfying $SU(2)$ commutation relations: 
\beq
[M_m,M_n] = i\sum_k \epsilon_{mnk} M_k . 
\label{s.53}
\eeq
The transposes satisfy
\beq
[ M^t_m, M^t_n] = i \sum_k \epsilon_{mnk} M^t_k
\label{s.54}
\eeq
\beq
[ (-M^t_m),(- M^t_n)] = i \sum_k \epsilon_{mnk} (-M^t_k)
\label{s.55}
\eeq
which shows that the negative transpose of these matrices also
satisfy $SU(2)$ commutation relations.  This shows that all of the
spin generators satisfy $SU(2)$ commutation relations.

Generators for rotations in
Euclidean space-time planes are constructed the same way from
\beq
\langle x,s,\nu \vert  J^{0\hat{n}} \vert \psi_{e} \rangle =   
\sum_\nu \left (i \delta_{\mu \nu} (\mathbf{x} {\partial \over \partial x^0} -
x^0 {\partial \over \partial \mathbf{x}}) 
- \langle s, \mu \vert \hat{\mathbf{n}}\cdot \mathbf{S} \vert s,\nu \rangle 
\right ) \langle x,s,\nu \vert \psi_e \rangle
\label{s.56}
\eeq
\beq
\langle x,s,\nu \vert  J^{0\hat{n}} \vert \psi_{ed} \rangle =
\sum_\nu \left (i \delta_{\mu \nu}(\mathbf{x} {\partial \over \partial x^0} -
x^0 {\partial \over \partial \mathbf{x}}) +
\langle s, \nu \vert \hat{\mathbf{n}}\cdot \mathbf{S} \vert s,\mu \rangle
\right ) \langle x,s,\nu \vert \psi_{ed} \rangle
\label{s.57}
\eeq
\beq
\langle x,s,\nu \vert  J^{0\hat{n}} \vert \psi_{e*} \rangle =  
\sum_\nu \left ( i \delta_{\mu \nu}(\mathbf{x} {\partial \over \partial x^0} -
x^0 {\partial \over \partial \mathbf{x}}) -
 \langle s, \nu \vert \hat{\mathbf{n}}\cdot \mathbf{S} \vert s,\mu \rangle 
\right ) \langle x,s,\nu \vert \psi_{e*} \rangle
\label{s.58}
\eeq
\beq
\langle x,s,\nu \vert  J^{0\hat{n}} \vert \psi_{ed*} \rangle =  
\sum_\nu \left ( i \delta_{\mu \nu}(\mathbf{x} {\partial \over \partial x^0} -
x^0 {\partial \over \partial \mathbf{x}}) 
+ \langle s, \mu \vert \hat{\mathbf{n}}\cdot \mathbf{S} \vert s,\nu \rangle 
\right ) \langle x,s,\nu \vert \psi_{ed*} \rangle
\label{s.59}
\eeq
In order to construct the boost generators it is necessary to multiply these
expression by an additional factor of (-i)
\beq
\langle x,s,\nu \vert  \mathbf{K}  \vert \psi_{e} \rangle =  
\sum_\nu \left ( \delta_{\mu \nu}(\mathbf{x} {\partial \over \partial x^0} -
x^0 {\partial \over \partial \mathbf{x}}) +
i \langle s, \mu \vert \hat{\mathbf{n}}\cdot \mathbf{S} \vert s,\nu \rangle 
\right ) \langle x,s,\nu \vert \psi_e \rangle
\label{s.60}
\eeq
\beq
\langle x,s,\nu \vert  \mathbf{K}  \vert \psi_{ed} \rangle =    
\sum_\nu \left ( \delta_{\mu \nu}(\mathbf{x} {\partial \over \partial x^0} -
x^0 {\partial \over \partial \mathbf{x}}) -i
\langle s, \nu \vert \hat{\mathbf{n}}\cdot \mathbf{S} \vert s,\mu \rangle
\right ) \langle x,s,\nu \vert \psi_{ed} \rangle
\label{s.61}
\eeq
\beq
\langle x,s,\nu \vert  \mathbf{K}  \vert \psi_{e*} \rangle =    
\sum_\nu \left ( \delta_{\mu \nu}(\mathbf{x} {\partial \over \partial x^0} -
x^0 {\partial \over \partial \mathbf{x}}) +i
 \langle s, \nu \vert \hat{\mathbf{n}}\cdot \mathbf{S} \vert s,\mu \rangle 
\right ) \langle x,s,\nu \vert \psi_{e*} \rangle
\label{s.62}
\eeq
\beq
\langle x,s,\nu \vert  \mathbf{K}  \vert \psi_{ed*} \rangle =    
\sum_\nu \left ( \delta_{\mu \nu} (\mathbf{x} {\partial \over \partial x^0} -
x^0 {\partial \over \partial \mathbf{x}}) 
-i \langle s, \mu \vert \hat{\mathbf{n}}\cdot \mathbf{S} \vert s,\nu \rangle 
\right ) \langle x,s,\nu \vert \psi_{ed*} \rangle
\label{s.63}
\eeq
The continuous part of these expressions agree with (\ref{p.10}-\ref{p.11})
for spinless operators.  The relevant commutators involving
the spin parts of the boost generators in each of the
four representations are
\beq
[K_i, K_j ]_{spin} = [i S_i,i S_j]  = -i \sum_k \epsilon_{ijk} S_k  =
-i \sum_k \epsilon_{ijk} J_{k\,spin}
\label{s.65}
\eeq
\beq
    [K_i, K_j ]_{spin} = [-i S^t_i,-i S^t_j] = -i \sum_k \epsilon_{ijk} (-S^t_k)
     =
    -i \sum_k \epsilon_{ijk} J_{k\,spin}
\label{s.65}
\eeq
\beq
[K_i, K_j ]_{spin} = [i S^t_i,i S^t_j] = -i \sum_k \epsilon_{ijk} (-S^t_k)
 =
    -i \sum_k \epsilon_{ijk} J_{k\,spin}
\label{s.66}
\eeq
\beq
[K_i, K_j ]_{spin} = [-i S_i,-i S_j]  = -i \sum_k \epsilon_{ijk} S_k 
 =
    -i \sum_k \epsilon_{ijk} J_{k\,spin}
\label{s.67}
\eeq
\beq
[K_i, S_j ]_{spin} = [i S_i, S_j]  = i \sum_k \epsilon_{ijk} (iS_k) = i \sum_k\epsilon_{ijk}  K_{k\,spin}
\label{s.68}
\eeq
\beq
[K_i, S_j ]_{spin} = [-iS_i^t, -S^t_j]  =  i\epsilon_{ijk}(-iS^t_k) = i\sum_k \epsilon_{ijk}  K_{k\,spin}
\label{s.69}
\eeq
\beq
[K_i, S_j ]_{spin} = [i S^t_i, -S^t_j]  =  i\epsilon_{ijk}  iS^t_k = i \sum_k\epsilon_{ijk}  K_{k\,spin}
\label{s.70}
\eeq
\beq
[K_i, S_j ]_{spin} = [-i S_i,S_j]  = i \epsilon_{ijk} (-iS_k) = \sum_k \epsilon_{ijk}  K_{k\,spin}
\label{s.71}
\eeq
where the spin generators in (\ref{s.65},\ref{s.66},\ref{s.69}) and
(\ref{s.70}) are (-) the transposes of the matrices satisfying $SU(2)$
commutation relations, which were shown in (\ref{s.53}-\ref{s.55}) to
satisfy $SU(2)$ commutation relations.  It follows that the
expressions (\ref{s.49}-\ref{s.52}) and (\ref{s.60}-\ref{s.63}) for the
Lorentz generators in each of the four spinor representations satisfy the
Poincar\'e commutation relations.

The hermiticity of these generators follows from the expressions
(\ref{s.36}-\ref{s.39}) and (\ref{s.40}-\ref{s.43}).
Each of equations (\ref{s.36}-\ref{s.39}) has the form
\beq
\langle U^{\dagger}(\lambda)\psi \vert \phi \rangle = 
\langle \psi \vert U(\lambda) \vert \phi \rangle 
\label{s.72}
\eeq
so the rotation operators, which are generators of
unitary one-parameter groups \cite{riesz} are self-adjoint in the
Hilbert spaces with inner products (\ref{s.12}-\ref{s.15}).

For the boost generators hermiticity follows from
(\ref{s.40}-\ref{s.43}).  In this case all of
these equations have the form
\beq
\langle T(\lambda)\psi \vert \phi \rangle = 
\langle \psi \vert T(\lambda) \vert \phi \rangle . 
\label{s.73}
\eeq
In these cases $T(\lambda)$ is Hermitian, but the generators are
constructed by multiplying the $\lambda$ derivative $1=(i)(-i)$
rather than $i$, resulting in Hermitian operators.

The self-adjointness of the Hamiltonian and boost generators is discussed
in the appendix.

In these covariant representations the spin does not enter in
the Hamiltonian or the linear momentum operators.   These operators
all commute with the spin operators and commutators with these
operators follow from the scalar case. 

The main result of this section is the expressions
(\ref{s.56}-\ref{s.63})
for the Poincar\'e generators.  These operators are formally
Hermitian on the different representations of the Euclidean
Hilbert space and they satisfy the Poincare commutation relations with
the translations generators (\ref{p.10}).

The construction in this section is limited to a description of
a particle of mass $m$ and spin $s$.

\section{Systems of Free Particles}

The Hilbert space to systems of free particles is the
direct sum of tensor product of single particle
Hilbert spaces.

Vectors in Euclidean Hilbert space for a system $N$ non-interacting
particles are represented by functions of the form
\beq
\psi (x_1, \mu_1, x_2, \mu_2 \cdots x_N, \mu_N)
\label{sys.1}
\eeq
with support for positive Euclidean time.  What is needed is that
the supports are non-overlapping and for positive time.
The indices $\mu_i$ are
$SU(2)\times SU(2)$ spinor indices.

The Hilbert space inner product is
\[
\langle \psi \vert \phi \rangle =
\sum \int d^{4N}x d^{4N}y
\psi^* ( x_1, \mu_1, x_2, \mu_2 \cdots,  x_N, \mu_N)
\prod_{n=1}^N S(\Theta x_n -y_n ,\mu_n,\nu_n)
\times
\]
\beq
\phi (y_1, \nu_1, y_2, \nu_2 \cdots y_N, \nu_N)
\label{sys.2}
\eeq
This is reflection positive since each of the $S(x_n -y_n ,\mu_n,\nu_n)$
is reflection positive and given explicitly by (\ref{s.8}-\ref{s.11}).
This is simply an
$N$-fold tensor product single-particle Hilbert spaces.

When the particles are identical the initial and final states can be
symmetrized or anti-symmetrized as appropriate.  Both of these
operations commute with the Euclidean time reflection and
consequently preserve the reflection positivity.  As long as
the Euclidean time supports are non-overlapping they can be re-ordered
so the relative times are positive. The
$\Theta x_n^0 - y_m^0$ will
always be negative.

Unlike the field theory case, the kernels are not assumed to be
completely symmetric (anti-symmetric) which leads to locality.
The Poincar\'e generators are sums of single particle generators

\section{Dynamics}

In Lorentz and Euclidean covariant representations of relativistic
quantum mechanics the dynamics enters through a kernel.
The reflection positivity constraint on Euclidean kernels is less
restrictive in the particle case than it is in the local field theory
case.  In a local field theory there is one $N$-point kernel for any
combination of $M$ initial degrees of freedom and $K$ final degrees of
freedom for $N=M+K$.  When locality is not required there can be
different reflection positive kernels for each combination of $M$
initial degrees of freedom and $K$ final degrees of freedom that add
up to $N$. This is a weaker form of reflection positivity.

For the purpose of making models it is desirable understand the
structure of model reflection positive kernels or to verify that model
kernels are reflection positive.

This section discusses the structure of reflection positive kernels.
This is illustrated by considering the example of a four-point
function, however the method can be applied to more general kernels.
Four-point functions have a cluster decomposition as the sum of
products of two-point functions and a connected four-point
function. The dynamics appears in the connected part of the four-point function.
Reflection positivity of two-point functions was demonstrated in
section IV.  This also true for products of these kernels.
A sufficient condition for the dynamical four-point function to
be reflection positive is that the connected part of the four-point
function is reflection positive.

The method used to construct reflection positive two-point functions
is used to examine the structure of reflection positive
four-point functions.  This construction is performed in two steps.
Lorentz covariant kernels are defined as vacuum expectation values of
formal Lorentz covariant fields and their adjoints.  Complete sets of
Poincar\'e irreducible states are inserted between the fields.  These
states are assumed to be positive mass-positive energy states.  These
are replaced by equivalent complete sets of Lorentz covariant
intermediate states.  This results in a decomposition of the kernel in
terms of Lorentz covariant matrix elements of Lorentz covariant
fields.  The spin structure of these matrix elements follows from the
covariance.  Analytic properties of covariant matrix elements that are
sufficient to construct an equivalent reflection positive Euclidean
kernel are identified.

In section IV right and left handed representations were treated
separately.  This section considers the general case of products of
these representations.  This is relevant for four vectors which
transform as a product of $j=1/2$ right and left handed
representations.  In order to distinguish the right and left handed
degrees of freedom, left handed spin degrees of freedom appear with a
dot superscript, $\dot j, \dot \mu$.

The analysis begins by considering fields 
\beq
\phi^{j \dot j}_{\mu \dot\mu}(x).
\label{dyn.1}
\eeq
that transform covariantly
under $SL(2,\mathbb{C})$: 
\beq
U(A) \phi^{j \dot{j}}_{\mu \dot\mu}(0) U^{\dagger} (A) =
\sum \phi^{j \dot j}_{\nu \dot\nu} (\Lambda x)
D^{j}_{\nu \mu}[A]  D^{\dot j}_{\dot \nu \dot\mu}[(A^{\dagger})^{-1}]
\label{dyn.2}
\eeq
and
\beq
U(A) \phi^{j \dot j \dagger}_{\mu \dot\mu}(x) U^{\dagger} (A) =
\sum \phi^{j \dot j \dagger}_{\nu \dot\nu} (\Lambda x)
D^{j'}_{\nu \mu}[A^*]  D^{\dot j}_{\dot \nu \dot\mu}[(A^{t})^{-1}] .
\label{dyn.3}
\eeq
Locality is not assumed.
Next consider the vacuum expectation value of the product
of two such fields and
their adjoints:
\[
\langle 0 \vert 
\phi^{j_2 \dot j_2 \dagger}_{\mu_2 \dot\mu_2}(x_2)
\phi^{j_1 \dot j_1 \dagger}_{\mu_1 \dot\mu_1}(x_1)
\phi^{j_1 \dot j_1}_{\nu_1 \dot\nu_1}(y_1)
\phi^{j_2 \dot j_2}_{\nu_2 \dot\nu_2}(y_2)
\vert 0 \rangle :=
\]
\beq
W_{2:2} (x_2,\mu_2, \dot\mu_2,x_1 ,\mu_1, \dot\mu_1 ;
y_1,\nu_1, \dot\nu_1,y_2 ,\nu_2, \dot\nu_2 ).
\label{dyn.4}
\eeq

This kernel is Lorentz covariant and 
{\it manifestly positive} since it has the from
\beq
\langle 0 \vert O^{\dagger} O \vert 0 \rangle . 
\label{dyn.5}
\eeq
This is referred to as a quasi-Wightman function. 
The next step is to insert complete sets of Poincar\'e irreducible
intermediate states between the fields.  Vacuum intermediate states do
not appear in the truncated part of the kernel.  The states that
appear are assumed to be positive mass intermediate states.

This results in a decomposition of (\ref{dyn.4}) the form
\[
W_{2:2} (x_2,\mu_2, \dot\mu_2,x_1 ,\mu_1, \dot\mu_1 ; y_1,\nu_1,
\dot\nu_1,y_2 ,\nu_2, \dot\nu_2 )  =
\]
\[
\sum \int 
\langle 0 \vert 
\phi^{j_2 \dot j_2}_{\mu_2 \dot\mu_2}(x_2)^{\dagger}
\vert p_a, \mu_a \rangle d\mathbf{p}_a
\langle  p_a, \mu_a \vert
\phi^{j_1 \dot j_1}_{\mu_1 \dot\mu_1}(x_1)^{\dagger}
\vert p_b, \mu_b \rangle d\mathbf{p}_b \times
\]
\beq
\langle  p_b, \mu_b \vert
\phi^{j_1 \dot j_1}_{\nu_1 \dot\nu_1}(y_1)
\vert p_c, \mu_c \rangle d\mathbf{p}_c
\langle  p_c, \mu_c \vert
\phi^{j_2 \dot j_2}_{\nu_2 \dot\nu_2}(y_2)
\vert 0 \rangle 
\label{dyn.6}
\eeq
where invariant quantum numbers have been suppressed.

To take advantage of the Lorentz covariance of the fields, the
intermediate states are replaced by equivalent Lorentz covariant
intermediate states as was done in section IV.  As was mentioned in
section IV, the Lorentz covariant states can be represented by right
or left handed representations.  In this application the Poincar\'e
covariant states are decomposed into products of spin states which are
transformed to products of right and left-handed Lorentz covariant
states.  This is done by decomposing the Poincar\'e irreducible
intermediate spins states into tensor products using $SU(2)$
Clebsch-Gordan coefficients.  In transforming to the Lorentz covariant
representation one factor is put in a right handed representation and
the other in a left-handed representation.

The resulting mixed Lorentz covariant states are defined by
\[
\vert (m,j)\mathbf {p} ,j_1,\dot j_2;\mu_1, \dot\mu_2 \rangle_{cov} :=
\]
\beq
\sum \vert (m,j) \mathbf{p} ,\mu \rangle \sqrt{\omega_m (\mathbf{p})}
\langle j, \mu \vert j_1, \nu_1, \dot j_2, \dot\nu_2 \rangle
D^{j_1}_{\nu_1 \mu_1} [B^{-1}_c (p)]
D^{\dot j_2}_{\dot \nu_2 \dot \mu_2} [B^{\dagger}_c (p)].
\label{dyn.7}
\eeq
These states transform covariantly
\[
U(A) \vert (m,j)\mathbf {p},j_1,\dot j_2;\mu_1, \dot\mu_2 \rangle_{cov} :=
\]
\beq
\sum \vert (m,j) \pmb{\Lambda}{p} ,j_1,\dot j_2;\nu_1, \dot\nu_2 \rangle_{cov}
D^{j_1}_{\nu_1 \mu_1} [A]
D^{\dot j_2}_{\dot \nu_2 \dot \mu_2} [(A^{\dagger})^{-1}].
\label{dyn.8}
\eeq
The identity can be expressed in terms of these states as
\[
I =\int \sum
\vert (m,j) \mathbf {p},\mu \rangle  d\mathbf{p}
\langle (m,j) \mathbf {p},\mu \rangle  =
\]
\beq
\sum
\int 
\vert (m,j)\mathbf {p},j_1,\dot j_2;\nu_1, \dot\nu_2 \rangle_{cov} 
{d\mathbf{p}\over \omega_m(\mathbf{p})}
D^{j_1}_{\nu_1 \mu_1} [p \cdot \sigma]
D^{\dot j_2}_{\dot\nu_2 \dot\mu_2} [\Pi p \cdot \sigma]
_{cov}\langle (m,j)\mathbf {p},j_1,\dot j_2;\mu_1, \dot\mu_2 \vert .
\label{dyn.9}
\eeq
The choice of how to break up the intermediate Poincar\'e covariant states
into right and left handed Lorentz covariant states is determined by
the spin structure of the fields.  

Replacing the Poincar\'e covariant intermediate states by
the corresponding mixed Lorentz covariant intermediate states in
(\ref{dyn.6}) gives
\[
W_{2:2} (x_2,\mu_2, \dot\mu_2,x_1 ,\mu_1, \dot\mu_1 ; y_1,\nu_1,
\dot\nu_1,y_2 ,\nu_2, \dot\nu_2 )  =
\]
\[
\sum \int 
\langle 0 \vert 
\phi^{j_2 \dot j_2 \dagger}_{\mu_2 \dot\mu_2}(x_2) 
\vert p_a, j_{a1},\dot{j}_{a2}; \mu_{a1}, \dot\mu_{a2} \rangle_{cov}
{d\mathbf{p}_a
\over \omega_{m_a}(\mathbf{p}_a)}
D^{j_{a1}}_{\mu_{a1} \nu_{a1}} [p \cdot \sigma]
D^{\dot{j}_{a2}}_{\dot\mu_{a2} \dot\nu_{a2}} [\Pi p \cdot \sigma] \times
\]
\[
{}_{cov}\langle p_a,j_{a1},\dot{j}_{a2}; \nu_{a1}, \dot\nu_{a2} \vert
\phi^{j_1 \dot j_1 \dagger}_{\mu_1 \dot\mu_1}(x_1)
\vert p_b,j_{b1},\dot{j}_{b2}; \mu_{b1}, \dot\mu_{b2}
\rangle_{cov} {d\mathbf{p}_b
\over \omega_{m_b}(\mathbf{p}_b)}
D^{j_{b1}}_{\mu_{b1} \nu_{b1}} [p \cdot \sigma]
D^{\dot{j}_{b2}}_{\dot\mu_{b2} \dot\nu_{b2}} [\Pi p \cdot \sigma]
\times \]
\[
{}_{cov}\langle p_b,j_{b1},\dot{j}_{b2}; \nu_{b1}, \dot\nu_{b2} \vert
\phi^{j_1 \dot j_1}_{\nu_1 \dot\nu_1}(y_1)
\vert p_c,j_{c1},\dot{j}_{c2}; \mu_{c1}, \dot\mu_{c2}
\rangle_{cov} {d\mathbf{p}_c
\over \omega_{m_c}(\mathbf{p}_c)}
D^{j_{c1}}_{\mu_{c1} \nu_{c1}} [p \cdot \sigma]
D^{\dot{j}_{c2}}_{\dot\mu_{c2} \dot\nu_{c2}} [\Pi p \cdot \sigma] \times
\]
\beq
{}_{cov}\langle p_c,j_{c1},\dot{j}_{c2}; \nu_{c1}, \dot\nu_{c2} \vert
\phi^{j_2' \dot j_2}_{\nu_2' \dot\nu_2}(y_2)
\vert 0 \rangle . 
\label{dyn.10}
\eeq
This expression contains four Lorentz covariant matrix elements of Lorentz
covariant field operators.

Translational covariance can be used to remove the space-time dependence from
each of these matrix elements
\beq
\langle 0 \vert 
\phi^{j_2 \dot j_2\dagger}_{\mu_2 \dot\mu_2}(x_2)
\vert p_a,j_{a1},\dot{j}_{a2}; \mu_{a1},\dot\mu_{a2} \rangle_{cov} =
e^{ip_a \cdot x_2}
\langle 0 \vert 
\phi^{j_2 \dot j_2 \dagger}_{\mu_2 \dot\mu_2}(0)
\vert p_a,j_{a1},\dot{j}_{a2}; \mu_{a1},\dot\mu_{a2} \rangle_{cov}
\label{dyn.11}
\eeq
\[
{}_{cov}\langle p_a,j_{a1},\dot{j}_{a2}; \nu_{a1}, \dot\nu_{a2} \vert
\phi^{j_1 \dot j_1 \dagger }_{\mu_1 \dot\mu_1}(x_1)
\vert p_b,j_{b1},\dot{j}_{b2}; \mu_{b1}, \dot\mu_{b2}
\rangle_{cov}
=
\]
\beq
e^{i (p_b-p_a)\cdot x_1}
{}_{cov}\langle p_a,j_{a1},\dot{j}_{a2}; \nu_{a1},\dot\nu_{a2} \vert
\phi^{j_1 \dot j_1}_{\mu_1  \dot\mu_1}(0)^{\dagger}
\vert p_b,j_{b1},\dot{j}_{b2}; \mu_{b1}, \dot\mu_{b2}
\rangle_{cov}
\label{dyn.12}
\eeq
\[
{}_{cov}\langle p_b,j_{b1},\dot{j}_{b2}; \nu_{b1}, \dot\nu_{b2} \vert
\phi^{j_1 \dot j_1}_{\nu_1 \dot\nu_1}(y_1)
\vert p_c,j_{c1},\dot{j}_{c2}; \mu_{c1}, \dot\mu_{c2}
\rangle_{cov}=
\]
\beq
e^{i (p_c-p_b)\cdot y_1}
{}_{cov}\langle p_b,j_{b1},\dot{j}_{b2}; \nu_{b1}, \dot\nu_{b2} \vert
\phi^{j_1 \dot j_1}_{\nu_1 \dot\nu_1}(0)
\vert p_c,,j_{c1},\dot{j}_{c2}; \mu_{c1}, \dot\mu_{c2}
\rangle_{cov}
\label{dyn.13}
\eeq
\beq
{}_{cov}\langle p_c,j_{c1},\dot{j}_{c2}; \nu_{c1}, \dot\nu_{c2} \vert
\phi^{j_2 \dot j_2}_{\nu_2 \dot\nu_2}(y_2)
\vert 0 \rangle =
e^{-i p_c \cdot y_2}
{}_{cov}\langle p_c,j_{c1},\dot{j}_{c2}; \nu_{c1}, \dot\nu_{c2} \vert
\phi^{j_2 \dot j_2}_{\nu_2 \dot\nu_2}(0)
\vert 0 \rangle .
\label{dyn.14}
\eeq
The Lorentz covariance properties of these matrix
elements with the space-time coordinate set to 0 are
\[
\langle 0 \vert 
\phi^{j_2 \dot j_2\dagger}_{\mu_2 \dot\mu_2}(0)
\vert p_a,j_{a1},\dot{j}_{a2}; \mu_{a1}, \dot\mu_{a2} \rangle_{cov} =
\]
\beq
\sum \langle 0 \vert 
\phi^{j_2 \dot j_2\dagger}_{\nu_2 \dot\nu_2}(0)
\vert \Lambda p_a,j_{a1},\dot{j}_{a2}; \nu_{a1} ,\dot\nu_{a2} \rangle_{cov}
D^{j_2}_{\nu_2 \mu_2}[A^*]D^{\dot j_2}_{\dot\nu_2 \dot\mu_2}[(A^t)^{-1}]
D^{j_{a1}}_{\nu_{a1} \mu_{a1}}[A]D^{\dot j_{a2}}_{\dot\nu_{a2} \dot\mu_{a2}}[(A^{\dagger})^{-1}]
\label{dyn.15}
\eeq
\[
{}_{cov}\langle p_a,j_{a1},\dot{j}_{a2}; \mu_{a1}, \dot\mu_{a2} \vert
\phi^{j_1 \dot j_1}_{\nu_1 \dot\nu_1}(0)^{\dagger}
\vert p_b,j_{b1},\dot{j}_{b2}; \mu_{b1}, \dot\mu_{b2}
\rangle_{cov} =
\]
\[
\sum {}_{cov}\langle \Lambda p_a,j_a,j_{a1},\dot{j}_{a2}; \nu_{a1}, \dot\nu_{a2} \vert
\phi^{j_1' \dot j_1}_{\mu'_1 \dot\mu_1}(0)^{\dagger}
\vert \Lambda p_b,j_b,j_{b1},\dot{j}_{b2}; \nu_{b1}, \dot\nu_{b2}
\rangle_{cov} \times
\]
\beq
D^{j_{a1}}_{\nu_{a1} \mu_{a1}}[A^*]D^{\dot j_{a2}}_{\dot\nu_{a2} \dot\mu_{a2}}[(A^t)^{-1}]
D^{j_1}_{\nu_1 \mu_1}[A^*]D^{\dot j_1}_{\dot\nu_1 \dot\mu_1}[(A^t)^{-1}]
D^{j_{b1}}_{\nu_{b1} \mu_{b1}}[A]D^{\dot j_{b2}}_{\dot\nu_{b2} \dot\mu_{b2}}[(A^{\dagger})^{-1}]
\label{dyn.16}
\eeq
\[
\sum
{}_{cov}\langle p_b,j_{b1},\dot{j}_{b2}; \nu_{b1}, \dot\nu_{b2} \vert
\phi^{j_1 \dot j_1}_{\nu_1 \dot\nu_1}(0)
\vert p_c,j_{c1},\dot{j}_{c2}; \mu_{c1}, \dot\mu_{c2}
\rangle_{cov} =
\]
\[
{}_{cov}\langle \Lambda p_b,j_{b1},\dot{j}_{b2}; \nu_{b1} ,\dot\nu_{b2} \vert
\phi^{j_1 \dot j_1}_{\nu_1 \dot\nu_1}(0)
\vert \Lambda p_c,j_{c1},\dot{j}_{c2}; \nu_{c1}, \dot\nu_{c2}
\rangle_{cov} \times
\]
\beq
D^{j_{b2}}_{\nu_{b2}' \mu_{b2}}[A^*]D^{\dot j_{b2}}_{\dot\nu_{b2} \dot\mu_{b2}}[(A^t)^{-1}]
D^{j_1}_{\nu_1 \mu_1}[A]D^{\dot j_1}_{\dot\nu_1 \dot\mu_1}[(A^{\dagger})^{-1}]
D^{j_{c1}}_{\nu_{c1} \mu_{c1}}[A]D^{\dot j_{c2}}_{\dot\nu_{c2} \dot\mu_{2}}[(A^{\dagger})^{-1}]
\label{dyn.17}
\eeq
\[
\sum
{}_{cov}\langle p_c,j_{c1},\dot{j}_{c2}; \nu_{c1}, \dot\nu_{b2} \vert
\phi^{j_2 \dot j_2}_{\nu_2 \dot\nu_2}(0)
\vert 0 \rangle =
\]
\beq
{}_{cov}\langle \Lambda p_c,j_{c1},\dot{j}_{c2}; \nu_{c1},
\dot\nu_{b2} \vert \phi^{j_2 \dot j_2}_{\nu_2 \dot\nu_2}(0) \vert 0
\rangle D^{j_{c2}}_{\nu_{c2} \mu_{c2}}[A^*]D^{\dot
j_{c2}}_{\dot\nu_{c2} \dot\mu_{c2}}[(A^t)^{-1}] D^{j_2}_{\nu_2\mu_2}
[A]D^{\dot j_2}_{\dot\nu_2 \dot\mu_2}[(A^{\dagger})^{-1}].
\label{dyn.18}
\eeq
In (\ref{dyn.16}-\ref{dyn.17}) $SU(2)$ Clebsch-Gordan coefficients
can be used to replace
\beq
D^{j_{a1}}_{\nu_{a1} \mu_{a1}}[A^*]D^{\dot j_{a2}}_{\dot\nu_{a2} \dot\mu_{a2}}[(A^t)^{-1}]
D^{j_1}_{\nu_1 \mu_1}[A^*]D^{\dot j_1}_{\dot\nu_1 \dot\mu_1}[(A^t)^{-1}]
\label{dyn.19}
\eeq
and
\beq
D^{j_{1'}}_{\nu_{1'} \mu_{1'}}[A]D^{\dot j_1}_{\dot\nu_1 \dot\mu_1}[(A^{\dagger})^{-1}]
D^{j_{b1}}_{\nu_{b1} \mu_{b1}}[A]D^{\dot j_{b2}}_{\dot\nu_{b2} \dot\mu_{b2}}[(A^{\dagger})^{-1}]\label{dyn.20}
\eeq
by
\beq
\sum
\langle j_{a1}, \nu_{a1}, j_{1}, \nu_{1} \vert j, \nu \rangle 
D^{j}_{\nu \mu}[A^*] \langle j, \mu, \vert j_{a1}, \mu_{a1}, j_1, \mu_1 \rangle
\langle \dot j_{a2}, \dot \nu_{a2}, \dot j_1, \dot \nu_1 \dot \vert \dot j,
\dot \nu \rangle 
D^{j}_{\dot \nu \dot\mu}[(A^t)^{-1}] \langle \dot j, \dot \mu, \vert \dot j_{a2}, \dot \mu_{a2}, \dot j_1, \dot \mu_1 \rangle
\label{dyn.21}
\eeq
and
\beq
\sum
\langle j_{1}, \nu_{1}, j_{b1}, \nu_{b1} \vert j, \nu \rangle 
D^{j}_{\nu \mu}[A] \langle j, \mu, \vert j_{1}, \mu_{1}, j_{b1}, \mu_{b1} \rangle
\langle \dot j_{1}, \dot \nu_{1}, \dot j_{b2}, \dot \nu_{b2} \dot \vert \dot j,
\dot \nu \rangle 
D^{j}_{\dot \nu \dot\mu}[(A^\dagger)^{-1}] \langle \dot j, \dot \mu, \vert \dot j_{1}, \dot \mu_{1}, \dot j_{2b}, \dot \mu_{2b} \rangle .
\label{dyn.22}
\eeq
After these replacements the spin dependence of
(\ref{dyn.16}) and (\ref{dyn.17}) 
has
the same structure as (\ref{dyn.15}) and (\ref{dyn.18}):
\beq
D^{j_{2}}_{\nu_{2'} \mu_{2}}[A^*]
D^{j_{a1}}_{\nu_{a1} \mu_{a1}}[A]
D^{\dot j_2}_{\dot\nu_2 \dot\mu_2}[(A^t)^{-1}]
D^{\dot j_{a2}}_{\dot\nu_{a2} \dot\mu_{a2}}[(A^{\dagger})^{-1}]
\label{dyn.23}
\eeq
and
\beq
D^{j_{c1}}_{\nu_{c1} \mu_{c1}}[A^*]
D^{j_2}_{\nu_{2} \mu_{2}}[A]
D^{\dot j_{c2}}_{\dot\nu_{c2} \dot\mu_{c2}}[(A^t)^{-1}]
D^{\dot j_2}_{\dot\nu_2 \dot\mu_2}[(A^{\dagger})^{-1}].
\label{dyn.24}
\eeq
The spins appearing in the fields are properties of the kernel.
The spins appearing in the intermediate states
are determined by the spin of the fields.
For example the spin of the field applied to the vacuum
fixes the spins of the first set of intermediate states. These states,
along with the spin of the second field fix the allowed spins in the
next set of intermediate, etc.

The building blocks of each of these covariant matrix elements are the
four momenta and the covariant spinors, $\sigma_{\mu},\sigma_{\mu}^*$
$\sigma_2 \sigma_{\mu}\sigma_2 ,\sigma_2\sigma_{\mu}^*\sigma_2$.
Functions of these quantities that
that have the transformation properties (\ref{dyn.23}-\ref{dyn.24})
follow from (\ref{b.8}):
\beq
D^{\dot j}_{\dot \mu \dot \nu}[p\cdot \sigma]
\qquad \dot j = \dot j_2 = \dot j_{a_2}
\qquad \dot j= \dot j_{c2} = \dot j_2 \qquad
D^{\dot j}_{\dot \mu \dot \nu}[p\cdot \sigma] =\sum
D^{\dot j}_{\dot \mu \dot \alpha}[\Lambda p\cdot \sigma]
D^{\dot j}_{\dot \alpha \dot \beta}[(A^t)^{-1}] D^{\dot j}_{\dot \beta \dot \nu}[(A^{\dagger})^{-1}]
\label{dyn.25}
\eeq
\beq
D^j_{\mu \nu}[p\cdot \sigma_2 \sigma^* \sigma_2]
\qquad 
j= j_2=j_{a2} 
\qquad
j= j_{c1}= j_2
\qquad
D^{j}_{\mu  \nu}[p\cdot \sigma_2 \sigma^* \sigma_2] = \sum
D^{j}_{\mu  \alpha}[\Lambda p\cdot \sigma_2 \sigma^* \sigma_2]
D^{j}_{\alpha \beta}[A^*] D^{j}_{\beta \nu}[A] 
\label{dyn.26}
\eeq
for any four momentum $p^{\mu}$ appearing in the matrix element.
This leads to the following covariant expressions for each of the matrix
elements:
\[
\langle 0 \vert 
\phi^{j_2 \dot j_2}_{\mu_2 \dot\mu_2}(0)^{\dagger}
\vert p_a,j_{a1},\dot{j}_{a2}; \mu_{a1}, \dot\mu_{a2} \rangle_{cov} =
\]
\beq
\delta_{j_2 j_{a1}} \delta_{\dot j_2 \dot j_a2}
F(j_2,j_{a1},\dot{j}_2,\dot{j}_{a2},p_a^2, m_a)
D^{j_2}_{\mu_2 \mu_{a1}}[p_a \cdot \sigma_2 \sigma^* \sigma_2] 
D^{\dot j_2}_{\dot \mu_2 \dot \mu_{a1}} [p_a\cdot \sigma] 
\label{dyn.27}
\eeq

\[
{}_{cov}\langle p_a,j_{a1},\dot{j}_{a2}; \mu_{1a}, \dot\mu_{a2} \vert
\phi^{j_1' \dot j_1}_{\mu'_1 \dot\mu_1}(0)^{\dagger}
\vert p_b,j_b,j_{b1},\dot{j}_{b2}; \mu_{b1}, \dot\mu_{b2}
\rangle_{cov} =
\]

\beq
\sum F_{ij} ( \cdots ) \times
\langle j_{a1}, \nu_{a1} ,j_{1}, \nu_{1},\vert  j_{b1}, \nu_{b1}  \rangle
D^{j_{b1}}_{\nu_{b1} \mu_{b1}}[p_i \cdot \sigma_2 \sigma^* \sigma_2]
\langle \dot j_{1}, \dot\nu_{1}, \dot j_{a2}, \dot\mu_{a2} \vert  \dot
j_{b2}, \dot \nu_{b2} \rangle
D^{\dot j_{b2}}_{\dot\nu_{b2} \dot \mu_{b2}}[p_j \cdot\sigma]
\label{dyn.28}
\eeq
\[
{}_{cov}\langle p_b,j_{b1},\dot{j}_{b2}; \nu_{b1}, \dot\nu_{b2} \vert
\phi^{j_1 \dot j_1}_{\nu'_1 \dot\nu_1}(0)
\vert p_c,j_{c1},\dot{j}_{c2}; \mu_{c1}, \dot\mu_{c2}
\rangle_{cov} =
\]
\beq
\sum F_{ij} (\cdots )
\langle j_{b1}, \nu_{1b} \vert j_{1}, \mu_{1}, j_{c1}, \mu_{c1} \rangle
D^{j_{b1}}_{\mu_{b1} \nu_{b1}}[p_i \cdot \sigma_2 \sigma^* \sigma_2]
D^{\dot j_{b2}}_{\dot\mu_{b2} \dot \nu_{b2}}[p_j \cdot\sigma]
\langle \dot
j_{21}, \dot \nu_{b2}\vert
\dot j_{1}, \dot\nu_{1}, \dot j_{c2}, \dot\mu_{c2},  \rangle
\label{dyn.29}
\eeq
\[
{}_{cov}\langle p_c,j_{c1},\dot{j}_{c2}; \nu_{c1}, \dot\nu_{c2} \vert
\phi^{j_2 \dot j_2}_{\nu_2 \dot\nu_2}(0)
\vert p_c,j_{c1},\dot{j}_{c2}; \mu_{c1}, \dot\mu_{c2}
\rangle_{cov} =
\]
\beq
F(j_1,j_{c1},\dot{j}_1,\dot{j}_{c2},p_c^2, m_c)
\delta_{j_{c1} j_1}\delta_{\dot j_{c2} \dot j_1}
\sum D^{\dot j_{1}}_{\dot\mu_{c2} \dot\mu_{1}}[p_j \cdot\sigma]
D^{j_{1}}_{\mu_{c1} \nu_{1}}[p_i \cdot \sigma_2 \sigma^* \sigma_2]
\label{dyn.30}
\eeq
where the coefficient functions, $F(\cdots )$, are scalars.
In the Lorentz covariant expressions all of the energies are on
shell.  The Wigner functions, $D[p\cdot \sigma]$ and $D[\Pi p \cdot
\sigma]$,  are all polynomials in the components of $p$.  In order
to get the equivalence it is necessary to assume
the coefficient functions are analytic functions of the
momenta in the upper or lower half energy
planes. 

Then to pass to the Euclidean representation replace all on shell
Lorentz four vectors by Euclidean four vectors, replace all of the
matrices $\sigma_{\mu}$ by $\sigma_{e\mu}$, and replace the factor ${d
  \mathbf{p} \over \omega_m(p)}$ appearing in the intermediate state
sums by ${d^4p_e \over p_e^2 + m^2}$.

When the kernel is smeared with Euclidean test functions satisfying
the Euclidean time support condition, each of the Euclidean
intermediate state integrals over $p^0_{0}$ results in a pole at $-i
\omega_m(p)$.  The results is $\sigma_{e} \cdot p_e \to
\omega_m (\mathbf{p}) I +\pmb{\sigma}\cdot \mathbf{p}$ and $p_{ie}\cdots
  p_j{je} \to p_1 \cdot p_j$.  This gives the equivalent
  Poincar\'e covariant kernel provided the rest of the integrand is
  analytic in the right or left half $p_e^0$ plane.  The Wigner
  functions of $p\cdot \sigma$ and $p \cdot \sigma_2 \sigma^* \sigma_2
  $ are polynomials in the components of $p_e$, so they don't generate
  any new singularities.  What remains are the scalar coefficient functions,
  $F(\cdots )$
  which must be analytic in the right and left half $p_e^0$ planes
  and have support for positive masses.

  This general structure is not surprising.  It illustrates how
  reflection positivity in the Euclidean representation is related to
  positivity and the spectral condition in the Lorentz covariant
  representation.  This is of limited value, since for models it
  requires building in the observed mass spectrum.  On the other hand
  it shows that the spin structures do not introduce new singularities.

A standard method to construct 
the Euclidean four-point kernel is related to the
Euclidean Bethe-Salpeter kernel $K$ by the Euclidean
Bethe-Salpeter equation, which has the form
\beq
S_4 = S_0 + S_0 K S_4 .
\label{dyn.31}
\eeq
This can be iterated to get
\beq
S_4 = S_0 + S_0 K S_0 + S_0 K S_0 K S_0 +\cdots .
\label{dyn.32}
\eeq
The corresponding series for the connected part of the four point
function is 
\beq
S_4^c = S_0 K S_0 + S_0 K S_0 K S_0 +\cdots 
\label{dyn.33}
\eeq
which can be formally expressed as
\beq
S_4^c = S_0 T S_0 \qquad T = K +K S_0 K + K  S_0 K S_0 K  +\cdots .
\label{dyn.34}
\eeq
Each term in the series for $K$ has the same Euclidean covariance
property.  If the series converges the sum will not generate any new
singularities that are not already in $K$.  This suggest that if
$S_0 K S_0$ is reflection positive and the series converges then
$S_0 T S_0$ will be reflection positive.  When the series does not
converge, the solution of the Bethe-Salpeter equation could result in
new singularities that violate the spectral condition.

These observations suggest that suitable Bethe-Salpeter kernels $K$
should have the property that $S_0 K S_0$ is reflection positive,
however this condition alone is not sufficient to ensure reflection
positivity of the resulting connected four point function.
The condition that  $S_0 K S_0$ is reflection positive should
be good starting point for constructing model dynamical four point
function.   

For dynamical models, while the dynamics appears in the kernel,
the expression for the Poincar\'e generators are sums of the
generators for each degree of freedom.

\section{Self Adjointness} 

While the self-adjointness of the generators of ordinary rotations
follows from the unitarity of the one-parameter group of rotations on
the Hilbert spaces (\ref{s.12}-\ref{s.15}), this argument does not
apply to either the Hamiltonian or the boost generators.  In both
cases the operators were derived from the corresponding Euclidean
generators by multiplication by an imaginary constant.  The Euclidean
generators and corresponding Lorentz generators act on different
Hilbert space representations.  The problem is that the corresponding
finite Euclidean transformations can map functions with positive time
support to functions that violate this condition.

For the Hamiltonian this can be treated by only considering
translations in the positive Euclidean time direction.
These translations map functions with positive Euclidean time
support into functions with positive Euclidean time support.
Reflection positivity can be used to show that
translations in the positive Euclidean time direction define a
contractive Hermitian semigroup on the Hilbert space with the scalar product
(\ref{s.12}-\ref{s.15}).
The argument \cite{glimm} uses the Schwartz inequality on both
the physical and Euclidean Hilbert spaces.  One application of
the Schwartz inequality on the physical Hilbert space gives
\beq
\Vert \vert e^{-H x^0} \vert \phi \rangle \Vert
= 
\langle e^{-H x^0} \phi \vert e^{-H x^0} \vert \phi \rangle^{1/2}
=
\langle \phi \vert e^{-H 2x^0} \vert \phi \rangle^{1/2} \leq
\Vert \vert e^{-H 2x^0} \vert \phi \rangle \Vert^{1/2}
\Vert \vert \phi \rangle \Vert^{1/2} .
\label{sa.1}
\eeq
Repeating these steps $n$-times gives
\beq
\Vert \vert e^{-H x^0} \vert \phi \rangle \Vert \leq
\Vert \vert e^{-H 2^nx^0} \vert \phi \rangle \Vert^{1/2^n}
\Vert \vert \phi \rangle \Vert^{1-1/2^n} .
\label{sa.2}
\eeq
The quantity
\beq
\Vert \vert e^{-H 2^nx^0} \vert \phi \rangle \Vert \leq
\Vert \theta U_e(2^nx^0) \vert \phi \rangle \Vert_e <
\Vert \vert \phi \rangle \Vert_e <
\infty
\label{sa.3}
\eeq
is bounded by the Euclidean norm, $\Vert \cdot \Vert_e$, 
since $U_e(2^nx^0)$ is unitary and $\Vert \theta \Vert_e= 1$ on
that Hilbert space.
Since this is finite and independent of $n$, taking the limit as
$n \to \infty$ gives
\beq
\Vert \vert e^{-H x^0} \vert \phi \rangle \Vert \leq
\Vert \vert \phi \rangle \Vert .
\label{sa.4}
\eeq
It follows that positive Euclidean time translations define a
contractive Hermitian semigroup on the Hilbert spaces (\ref{s.12}-\ref{s.15}).  
The generator is a positive self-adjoint operator \cite{riesz}\cite{simon}.

Boosts present additional complications.  Even an infinitesimal
rotation in a Euclidean space time plane will map a general function
with positive Euclidean time support to one that violates this
condition.  The self-adjointness of the boost generator cannot be
demonstrated by showing that it defines a unitary one-parameter group
or contractive semigroup, however it turns out that rotations in
Euclidean space time planes, which are interpreted as boosts with
complex rapidity, define local symmetric semigroups
\cite{Klein:1981}\cite{Klein:1983} \cite{Frohlich:1983kp} on the
Hilbert spaces (\ref{s.12}-\ref{s.15}).  These have self-adjoint
generators, which are exactly the boost generators.

The conditions for a local symmetric semigroup 
\cite{Klein:1981} 
are
\begin{itemize}
\item[1.] For each $\theta \in [0 ,\theta_0]$, there is a linear subset
${\cal D}_{\theta}$ such that  ${\cal D}_{\theta_1}\supset  {\cal D}_{\theta_2}$
if $\theta_1< \theta_2$, and $\cup_{0<\theta<\theta_0}
{\cal D}_{\theta_2}$ is dense.

\item[2.] For each $\theta \in [0,\theta_0]$, $E(\theta)$ is a linear operator
  on the Hilbert space with domain ${\cal D}_{\theta}$

\item[3.] $E(0)=I$,
  $E(\theta_1):{\cal D}_{\theta_2}\to  {\cal D}_{\theta_2-\theta_1}$,
  and $E(\theta_1)E(\theta_2)= E(\theta_1+\theta_2)$ on
  ${\cal D}_{\theta_1+\theta_2}$ for $\theta_1,\theta_2,\theta_1+\theta_2 \in
  [0,\theta_0]$
  
\item[4.] $E(\theta)$ is Hermitian for $\theta \in [0,\theta_0]$ 
  
\item[5.] $E(\theta)$ is weakly continuous on $[0,\theta_0]$  

\end{itemize}

When these conditions are satisfied there is a unique self-adjoint
operator $K$ such that ${\cal D}_{\theta} \subset   {\cal D}_{e^{-K\theta}}$
and $E(\theta)$ is the restriction of $e^{-K\theta}$ to
${\cal D}_{\theta}$.

In this case $E(\theta)$ represents Euclidean space time rotations
considered as operators on the Hilbert space (\ref{s.12}-\ref{s.15}) 
restricted to domains that will be described below.

The domains are Schwartz functions with space Euclidean time support the
wedge shaped region defined by
\beq
\mathbf{x} \cdot \hat{\mathbf{n}} - {x^0_e \over \epsilon} + \epsilon <0
\label{sa.5}
\eeq
\beq
\mathbf{x} \cdot \hat{\mathbf{n}} + {x^0_e \over \epsilon} - \epsilon >0
\label{sa.6}
\eeq

The wedge shaped region becomes the positive Euclidean time
half plane in the limit
that $\epsilon\to 0$.  Schwartz functions with support on this
half plane are dense.  In addition,
if this domain is rotated  by an angle less than 
$\theta_\epsilon :=\pm \tan^{-1}(\epsilon)$, it
will still be contained in the positive Euclidean time
half plane.  Schwartz functions with support in
these wedge shaped regions can be constructed from
Schwartz functions that have support or positive Euclidean time
by multiplying the function by 
$g(x^0,\mathbf{x}\cdot \hat{\mathbf{n}},\epsilon)$
where 
\beq
g(x^0,\mathbf{x}\cdot \hat{\mathbf{n}},\epsilon)) =
h({x^0_e \over \epsilon} - \epsilon +\mathbf{x} \cdot \hat{\mathbf{n}})
h({x^0_e \over \epsilon} - \epsilon -\mathbf{x} \cdot \hat{\mathbf{n}}).
\label{sa.7}
\eeq
and
\beq
h (\lambda)  =
\left \{
\begin{array}{cc}
e^{- {1 \over (\lambda)^2}} & \lambda >0 \\
0 & \lambda \leq 0 \\  
\end{array} 
\right . .
\label{sa.8}
\eeq
is a  smoothed Heaviside function.
$g(x^0,\mathbf{x}\cdot \hat{\mathbf{n}},\epsilon))$ is a Schwartz function with
support in the wedge shaped region (\ref{sa.5}-\ref{sa.6}) that
approaches $1$ as $\epsilon (\theta)$ approaches 0.

The domain ${\cal D}_{\theta}$ is taken as the space of Schwartz functions with
positive time support multiplied by the function
$g(x^0,\mathbf{x}\cdot \hat{\mathbf{n}},\epsilon))$ where $\theta=\theta_\epsilon$.  The Euclidean space time rotations restricted to
these  domains have all of the properties of local symmetric semigroup.
It follows that the boost generators $\mathbf{K}$ are self-adjoint
on the physical Hilbert space.

\section{Summary and Conclusion} 

Relativistic formulations of quantum mechanics are useful for
understanding the short-distance properties of strongly interacting
systems.  The advantage is that they can be solved using the same
Hilbert space methods that are used in non-relativistic quantum
theories.  The challenges are formulating the models so isolated
subsystems are separately Poincar\'e invariant (cluster properties)
and relating the phenomenological interactions to QCD.  A Euclidean
approach provides a one way of addressing these challenges, while
creating a different set of challenges.  The Euclidean formulation of
relativistic quantum mechanics is motivated by the axioms of Euclidean
quantum field theory.  The axiom that leads to microscopic locality is
logically independent of the other axioms.  This investigation of
Euclidean formulations of relativistic quantum theories of particles
is motivated by the possibility of being able to satisfy all of the
axioms of relativistic quantum field theory without having to require
locality.  While microscopic locality is desirable, it is the source
of most of the difficulties of quantum field theory, and is difficult
to test experimentally.  In the Euclidean formulation cluster
properties can be easily satisfied, there is a natural relation to
quantum field theories, and the formalism is still a theory of linear
operators acting on a Hilbert space.

An appealing feature of the Euclidean representation is that the physical
Hilbert space and the infinitesimal generators of the Poincar\'e group
can be constructed without any need for an analytic continuation to
Minkowski space.

The new property of this representation is that the Hilbert space
inner product has a non-trivial kernel that is not manifestly
symmetric.  The requirement that Hilbert space inner product has a
positive norm is called reflection positivity, which constrains the
form of the kernel.  While being able to use standard Hilbert space
methods in the Euclidean representation has some advantages, these
methods get modified in unfamiliar ways when the inner product has a
non-trivial kernel.  Among the unfamiliar properties are
that self-adjoint operators have unfamiliar forms,
distributions, like delta functions, can become normalizable vectors,
and the Poincar\'e generators have a form that does not depend on the
interactions.  In addition, a deviation from the Euclidean formulation
of field theory is that a single $N$-point kernel can be replaced by
$N-1$, $K+M=N$ point kernels.

The primary purpose of this paper is to provide explicit expressions
for the Poincar\'e generators for particles of any spin in the
Euclidean representation.  While these formulas follow from the
definitions in a straightforward manner, explicit formulas are needed
for applications, especially for hadronic and nuclear physics
applications that can involve particles with high spins.  Since the
forms of these operators are unfamiliar, the commutation relations and
self-adjointness of each one is demonstrated explicitly. 

Generators and Euclidean kernels were derived by starting with
positive mass irreducible representations of the Poincar\'e group,
constructing equivalent Lorentz covariant representations, and using
these to construct Euclidean covariant representations.  This
automatically results in reflection positive irreducible
representations.  These results are general since any unitary
representation of the Poincar\'e group can be decomposed into a direct
integral of positive-mass positive-energy irreducible representations.
While this also applies to systems of particles, for systems it is
useful to replace the irreducible representation by products of single
particle irreducible representations.  Interactions require
introducing correlations in the $N$ free particle kernel.  While the
correlations preserve the covariance properties, the requirement that
they preserve reflection positivity is not automatic.  The formulas
for the Poincar\'e generators remain unchanged.

The structure of reflection positive kernels with arbitrary spin was
investigated.  On one hand the spin structures that result from
covariance do not impact the reflection positivity, however the
analytic properties of the coefficient functions must satisfy a
spectral condition and be analytic in the right and left half
Euclidean energy planes.  Ideally one would like to be able find
sufficient conditions on the input to Schwinger Dyson equations so the
solution generates reflection positive kernels.  For the simplest case
of the Bethe-Salpeter equation, while the equation preserves the spin
structure, the solution can introduce singularities that violate
reflection positivity.

In the Euclidean representation the dynamics appears in the Hilbert
space kernel.  While in principle Hilbert space methods can be used in
calculations, because of the kernel, applications favor different
methods of computation.  These have been discussed elsewhere,
\cite{Kopp:2011vv}\cite{Polyzou:2013nga}
\cite{Aiello:2015jgc}\cite{polyzou:2019}.  In addition, the triviality
of the Poincar\'e generators puts the burden of constructing dynamical
models on the structure of reflection positive kernels.

The authors would like to acknowledge Palle J{\o}rgensen for helpful discussions
on reflection positivity.

\section{Appendix}

For a number of applications it is necessary to compute
\[
{d^n \over dx^n} {K_1 (x) \over x} =
\]
\[
\sum_{m=0}^n {n! \over m! (n-m)!}
{d^m \over dx^m} K_1 (x) (-)^{n-m} {(n-m)!\over x^{n-m-1}}=
\]
\[
\sum_{m=0}^n {n! (-)^{n-m} \over m!x^{n-m+1} }
{d^m \over dx^m} K_1 (x) =
\]
\beq
{d^m \over dx^m} K_1 (x)  =
{1 \over 2^m}
(-)^{1-m}( K_{1-m}(x)  + {m! \over 1! (m-1)!} K_{1-m-2}(x)
{m! \over 2! (m-2)!} K_{1-m-4}(x) +\cdots +
K_{1+m}(x) 
)
\eeq
where
\beq
K_n(x) = K_{-n}(x)
\eeq
and
\beq
K_{n+1}(x) = K_{n-1}(x) + {2n \over x} K_n(x)
\eeq
\bibliography{master_bibfile}

\begin{thebibliography}{28}
\expandafter\ifx\csname natexlab\endcsname\relax\def\natexlab#1{#1}\fi
\expandafter\ifx\csname bibnamefont\endcsname\relax
  \def\bibnamefont#1{#1}\fi
\expandafter\ifx\csname bibfnamefont\endcsname\relax
  \def\bibfnamefont#1{#1}\fi
\expandafter\ifx\csname citenamefont\endcsname\relax
  \def\citenamefont#1{#1}\fi
\expandafter\ifx\csname url\endcsname\relax
  \def\url#1{\texttt{#1}}\fi
\expandafter\ifx\csname urlprefix\endcsname\relax\def\urlprefix{URL }\fi
\providecommand{\bibinfo}[2]{#2}
\providecommand{\eprint}[2][]{\url{#2}}

\bibitem[{\citenamefont{Wigner}(1939)}]{Wigner:1939cj}
\bibinfo{author}{\bibfnamefont{E.~P.} \bibnamefont{Wigner}},
  \bibinfo{journal}{Annals Math.} \textbf{\bibinfo{volume}{40}},
  \bibinfo{pages}{149} (\bibinfo{year}{1939}).

\bibitem[{\citenamefont{Bargmann}(1954)}]{Bargmann:1954gh}
\bibinfo{author}{\bibfnamefont{V.}~\bibnamefont{Bargmann}},
  \bibinfo{journal}{Annals Math.} \textbf{\bibinfo{volume}{59}},
  \bibinfo{pages}{1} (\bibinfo{year}{1954}).

\bibitem[{\citenamefont{Sokolov}(1977)}]{Sokolov:1977}
\bibinfo{author}{\bibfnamefont{S.~N.} \bibnamefont{Sokolov}},
  \bibinfo{journal}{Dokl. Akad. Nauk SSSR} \textbf{\bibinfo{volume}{233}},
  \bibinfo{pages}{575} (\bibinfo{year}{1977}).

\bibitem[{\citenamefont{Coester and Polyzou}(1982)}]{Coester:1982vt}
\bibinfo{author}{\bibfnamefont{F.}~\bibnamefont{Coester}} \bibnamefont{and}
  \bibinfo{author}{\bibfnamefont{W.~N.} \bibnamefont{Polyzou}},
  \bibinfo{journal}{Phys. Rev.} \textbf{\bibinfo{volume}{D26}},
  \bibinfo{pages}{1348} (\bibinfo{year}{1982}).

\bibitem[{\citenamefont{Keister and Polyzou}(1991)}]{Keister:1991sb}
\bibinfo{author}{\bibfnamefont{B.~D.} \bibnamefont{Keister}} \bibnamefont{and}
  \bibinfo{author}{\bibfnamefont{W.~N.} \bibnamefont{Polyzou}},
  \bibinfo{journal}{Adv. Nucl. Phys.} \textbf{\bibinfo{volume}{20}},
  \bibinfo{pages}{225} (\bibinfo{year}{1991}).

\bibitem[{\citenamefont{Schwinger}(1958)}]{Schwinger:pna}
\bibinfo{author}{\bibfnamefont{J.~S.} \bibnamefont{Schwinger}},
  \bibinfo{journal}{Proc. Natl. Acad. Sci. U. S.}
  \textbf{\bibinfo{volume}{44}}, \bibinfo{pages}{956} (\bibinfo{year}{1958}).

\bibitem[{\citenamefont{Schwinger}(1959)}]{Schwinger:1959zz}
\bibinfo{author}{\bibfnamefont{J.}~\bibnamefont{Schwinger}},
  \bibinfo{journal}{Phys. Rev.} \textbf{\bibinfo{volume}{115}},
  \bibinfo{pages}{721} (\bibinfo{year}{1959}).

\bibitem[{\citenamefont{Streater and Wightman}(1980)}]{Wightman:1980}
\bibinfo{author}{\bibfnamefont{R.~F.} \bibnamefont{Streater}} \bibnamefont{and}
  \bibinfo{author}{\bibfnamefont{A.~S.} \bibnamefont{Wightman}},
  \emph{\bibinfo{title}{PCT, Spin and Statistics, and All That}}
  (\bibinfo{publisher}{Princeton Landmarks in Physics}, \bibinfo{year}{1980}).

\bibitem[{\citenamefont{Jost}(1965)}]{jost}
\bibinfo{author}{\bibfnamefont{R.}~\bibnamefont{Jost}},
  \emph{\bibinfo{title}{The General Theory of Quantized Fields}}
  (\bibinfo{publisher}{AMS}, \bibinfo{year}{1965}).

\bibitem[{\citenamefont{Symanzik}(1966)}]{Symanzik:1966}
\bibinfo{author}{\bibfnamefont{K.}~\bibnamefont{Symanzik}},
  \bibinfo{journal}{J. Math. Phys.} \textbf{\bibinfo{volume}{7}},
  \bibinfo{pages}{510} (\bibinfo{year}{1966}).

\bibitem[{\citenamefont{Symanzik}(1968)}]{Symanzik:1968zz}
\bibinfo{author}{\bibfnamefont{K.}~\bibnamefont{Symanzik}},
  \bibinfo{journal}{Conf. Proc.} \textbf{\bibinfo{volume}{C680812}},
  \bibinfo{pages}{152} (\bibinfo{year}{1968}).

\bibitem[{\citenamefont{Nelson}(1973)}]{Nelson:1973}
\bibinfo{author}{\bibfnamefont{E.}~\bibnamefont{Nelson}}, \bibinfo{journal}{J.
  Funct. Anal.} \textbf{\bibinfo{volume}{12}}, \bibinfo{pages}{97}
  (\bibinfo{year}{1973}).

\bibitem[{\citenamefont{Osterwalder and Schrader}(1973)}]{Osterwalder:1973dx}
\bibinfo{author}{\bibfnamefont{K.}~\bibnamefont{Osterwalder}} \bibnamefont{and}
  \bibinfo{author}{\bibfnamefont{R.}~\bibnamefont{Schrader}},
  \bibinfo{journal}{Commun. Math. Phys.} \textbf{\bibinfo{volume}{31}},
  \bibinfo{pages}{83} (\bibinfo{year}{1973}).

\bibitem[{\citenamefont{Osterwalder and Schrader}(1975)}]{Osterwalder:1974tc}
\bibinfo{author}{\bibfnamefont{K.}~\bibnamefont{Osterwalder}} \bibnamefont{and}
  \bibinfo{author}{\bibfnamefont{R.}~\bibnamefont{Schrader}},
  \bibinfo{journal}{Commun. Math. Phys.} \textbf{\bibinfo{volume}{42}},
  \bibinfo{pages}{281} (\bibinfo{year}{1975}).

\bibitem[{\citenamefont{Kopp and Polyzou}(2012)}]{Kopp:2011vv}
\bibinfo{author}{\bibfnamefont{P.}~\bibnamefont{Kopp}} \bibnamefont{and}
  \bibinfo{author}{\bibfnamefont{W.}~\bibnamefont{Polyzou}},
  \bibinfo{journal}{Phys. Rev.} \textbf{\bibinfo{volume}{D85}},
  \bibinfo{pages}{016004} (\bibinfo{year}{2012}), \eprint{1106.4086}.

\bibitem[{\citenamefont{Polyzou}(2014)}]{Polyzou:2013nga}
\bibinfo{author}{\bibfnamefont{W.~N.} \bibnamefont{Polyzou}},
  \bibinfo{journal}{Phys. Rev.} \textbf{\bibinfo{volume}{D89}},
  \bibinfo{pages}{076008} (\bibinfo{year}{2014}), \eprint{1312.3585}.

\bibitem[{\citenamefont{Aiello and Polyzou}(2016)}]{Aiello:2015jgc}
\bibinfo{author}{\bibfnamefont{G.}~\bibnamefont{Aiello}} \bibnamefont{and}
  \bibinfo{author}{\bibfnamefont{W.}~\bibnamefont{Polyzou}},
  \bibinfo{journal}{Phys. Rev.} \textbf{\bibinfo{volume}{D93}},
  \bibinfo{pages}{056003} (\bibinfo{year}{2016}), \eprint{1512.03651}.

\bibitem[{\citenamefont{Polyzou}(2019)}]{polyzou:2019}
\bibinfo{author}{\bibfnamefont{W.~N.} \bibnamefont{Polyzou}},
  \bibinfo{journal}{Phys. Rev.} \textbf{\bibinfo{volume}{C99}},
  \bibinfo{pages}{025202} (\bibinfo{year}{2019}).

\bibitem[{\citenamefont{Rose}(1957)}]{rose}
\bibinfo{author}{\bibfnamefont{M.}~\bibnamefont{Rose}},
  \emph{\bibinfo{title}{Elementary Theory of Angular Momentum}}
  (\bibinfo{publisher}{Wiley}, \bibinfo{year}{1957}).

\bibitem[{\citenamefont{Bogoliubov and Shirkov}(1959)}]{bogoliubov}
\bibinfo{author}{\bibfnamefont{N.~N.} \bibnamefont{Bogoliubov}}
  \bibnamefont{and} \bibinfo{author}{\bibfnamefont{D.~V.}
  \bibnamefont{Shirkov}}, \emph{\bibinfo{title}{Introduction to the theory of
  quantized fields}} (\bibinfo{publisher}{Wiley-Interscience},
  \bibinfo{year}{1959}).

\bibitem[{\citenamefont{Wightman}(1960)}]{Wightman}
\bibinfo{author}{\bibfnamefont{A.~S.} \bibnamefont{Wightman}},
  \emph{\bibinfo{title}{{L'Invariance Dans La Mecanique Quantique
  Relativiste}}}, vol.~\bibinfo{volume}{7} (\bibinfo{publisher}{Hermann,
  Paris}, \bibinfo{year}{1960}).

\bibitem[{\citenamefont{Berestetskii et~al.}(1982)\citenamefont{Berestetskii,
  Lifshitz, and Pitaevskii}}]{berestetskii}
\bibinfo{author}{\bibfnamefont{V.~B.} \bibnamefont{Berestetskii}},
  \bibinfo{author}{\bibfnamefont{E.~M.} \bibnamefont{Lifshitz}},
  \bibnamefont{and} \bibinfo{author}{\bibfnamefont{L.~P.}
  \bibnamefont{Pitaevskii}}, \emph{\bibinfo{title}{Quantum Electrodynamics}}
  (\bibinfo{publisher}{Pergammon Press, Elmsford N.Y.}, \bibinfo{year}{1982}).

\bibitem[{\citenamefont{Riesz and Sz.Nagy}(1972)}]{riesz}
\bibinfo{author}{\bibfnamefont{F.}~\bibnamefont{Riesz}} \bibnamefont{and}
  \bibinfo{author}{\bibfnamefont{B.}~\bibnamefont{Sz.Nagy}},
  \emph{\bibinfo{title}{Functional Analysis}} (\bibinfo{publisher}{Ungar,
  N.Y.}, \bibinfo{year}{1972}).

\bibitem[{\citenamefont{Gilmm and Jaffe}(1981)}]{glimm}
\bibinfo{author}{\bibfnamefont{J.}~\bibnamefont{Gilmm}} \bibnamefont{and}
  \bibinfo{author}{\bibfnamefont{A.}~\bibnamefont{Jaffe}},
  \emph{\bibinfo{title}{Quantum Physics - A functional Integral Point of View}}
  (\bibinfo{publisher}{Springer}, \bibinfo{year}{1981}).

\bibitem[{\citenamefont{Reed and Simon}(1979)}]{simon}
\bibinfo{author}{\bibfnamefont{M.}~\bibnamefont{Reed}} \bibnamefont{and}
  \bibinfo{author}{\bibfnamefont{B.}~\bibnamefont{Simon}},
  \emph{\bibinfo{title}{Methods of Modern mathematical Physics}}, vol.
  \bibinfo{volume}{III Scattering Theory} (\bibinfo{publisher}{Academic Press},
  \bibinfo{year}{1979}).

\bibitem[{\citenamefont{Klein and L.}(1981)}]{Klein:1981}
\bibinfo{author}{\bibfnamefont{A.}~\bibnamefont{Klein}} \bibnamefont{and}
  \bibinfo{author}{\bibfnamefont{L.}~\bibnamefont{L.}}, \bibinfo{journal}{J.
  Functional Anal.} \textbf{\bibinfo{volume}{44}}, \bibinfo{pages}{121}
  (\bibinfo{year}{1981}).

\bibitem[{\citenamefont{Klein and L.}(1983)}]{Klein:1983}
\bibinfo{author}{\bibfnamefont{A.}~\bibnamefont{Klein}} \bibnamefont{and}
  \bibinfo{author}{\bibfnamefont{L.}~\bibnamefont{L.}}, \bibinfo{journal}{Comm.
  Math. Phys} \textbf{\bibinfo{volume}{87}}, \bibinfo{pages}{469}
  (\bibinfo{year}{1983}).

\bibitem[{\citenamefont{Frohlich et~al.}(1983)\citenamefont{Frohlich,
  Osterwalder, and Seiler}}]{Frohlich:1983kp}
\bibinfo{author}{\bibfnamefont{J.}~\bibnamefont{Frohlich}},
  \bibinfo{author}{\bibfnamefont{K.}~\bibnamefont{Osterwalder}},
  \bibnamefont{and} \bibinfo{author}{\bibfnamefont{E.}~\bibnamefont{Seiler}},
  \bibinfo{journal}{Annals Math.} \textbf{\bibinfo{volume}{118}},
  \bibinfo{pages}{461} (\bibinfo{year}{1983}).

\end{thebibliography}
\end{document}